\documentclass{aa}

\usepackage{cuted}
\usepackage{graphicx}
\usepackage{footnote}
\usepackage{booktabs}
\usepackage{placeins}
\usepackage{lscape}

\usepackage{dcolumn}

\newcolumntype{d}[1]{D{?}{\,\pm\,}{#1}}

\makesavenoteenv{tabular}
\makesavenoteenv{table}
\usepackage{multirow}

\usepackage{cuted}
\usepackage{caption}

\usepackage{enumitem}

\usepackage{txfonts}
\usepackage[dvipsnames]{xcolor}
\usepackage[colorlinks=true,pdfborder={0 0 0},linkcolor=Blue,citecolor=Blue,urlcolor=Green]{hyperref}

\newcommand{\loggf}[1]{$\log(gf)_\mathrm{#1}$}
\newcommand{\dloggf}[1]{$\Delta\log(gf)_\mathrm{#1}$}

\newcommand{\globin}{\texttt{globin}}
\newcommand{\FusionG}{Fusion$(\hat{G})$}
\newcommand{\FusionLG}{Fusion$(\hat{L}\oplus\hat{G})$}
\newcommand{\Gcoupled}{$\hat{G}_\mathrm{coupled}$}

\newcommand{\eq}[1]{Eq.~(\ref{#1})}
\newcommand{\fig}[1]{Fig.~\ref{#1}}
\newcommand{\sref}[1]{Sec.~\ref{#1}}
\newcommand{\tab}[1]{Tab.~\ref{#1}}

\begin{document}
   \title{Robustness of transition probabilities of spectral lines deduced from spatially resolved solar spectra}

   \author{D. Vukadinovi\'{c}\inst{1,2},
          H. N. Smitha\inst{1},
          A. Lagg\inst{1},
          S. K. Solanki\inst{1,3}
          and
          J. C. Trelles Arjona\inst{4,5}
          }

   \institute{Max Planck Institute for Solar System Research, Justus-von-Liebig Weg 3, G\"{o}ttingen\\
        \email{dusan.vukadinovic@uni-graz.at}
            \and
        Institute of Physics, University of Graz, Universit\"{a}tplatz 5, 8010, Graz, Austria
             \and
        School of Space Research, Kyung Hee University, Yongin, Gyeonggi 17104, Republic of Korea
            \and
        Instituto de Astrofísica de Canarias (IAC), Vía Láctea s/n, E-38205 La Laguna, Tenerife, Spain
            \and
        Universidad de La Laguna, Departamento de Astrofísica, E-38206 La Laguna, Tenerife, Spain
    }

   \date{Received month dd, yyyy; accepted month dd, yyy}

\authorrunning{Vukadinovi\'{c} et al.}
\titlerunning{Robustness of the astrophysically calibrated transition probabilities of atomic lines from solar spectra}

  \abstract
  {Atomic parameters are a key ingredient for the synthesis of spectral lines to model observed spectra and to extract meaningful physical information about the observed astrophysical object.}
  {The atomic line parameters can be determined with different approaches from observed solar and stellar spectra. We aim to use quiet-sun observations to test the reliability of the atomic parameter inference from observed spectra. Our focus here is on the transition probability of spectral lines, usually denoted as \loggf{}.}
  {We apply a spectropolarimetric inversion method to quiet-sun disc-centre observations at 1.56\,$\mu$m from the GRIS/GREGOR spectrograph. We use the coupled method implemented in the \globin{} inversion code, where atomic parameters are spatially coupled, and a single value of these parameters is determined for every spectral line over the entire field of view. The underlying assumption is that our line list is complete and that all lines satisfy local thermodynamic equilibrium. We carry out tests to determine the sensitivity of the inferred \loggf{} values to various atmospheric parameters and the data reduction fallibility.}
  {The inferred \loggf{} values are compared to those from a recent study where \loggf{} values were determined using the same observations but a different method. We find that our \loggf{} values are weaker by $0.023$\,dex, which is most probably caused by the constraint imposed on the \loggf{} parameter due to the spatial coupling, lifting the degeneracy between temperature and \loggf{}. We use the Markov chain Monte Carlo method to determine the uncertainty in the inferred \loggf{} values. For the strongest lines in the spectral region, the uncertainty in \loggf{} is $1-4$\,\%, increasing to $12$\,\% for weaker and blended lines.}
  {Our tests suggest that the \loggf{} values inferred using the coupled method from \globin{} are robust against any local contamination to the observed line profiles, such as spectral stray light, and return a reliable and unique set of \loggf{} values for the strongest lines irrespective of the chosen atmospheric model and observation dataset. This result is valid as long as the underlying assumptions of line formation under local thermodynamic equilibrium and the completeness of the line list are satisfied.}

   \maketitle
   \nolinenumbers
 
\section{Introduction}

The reliability of the atmospheric physical parameters inferred from observed spectra depends greatly on the adopted atomic parameters used in the computations to reproduce the observed spectral lines. The relevant atomic parameters for modelling spectral lines are the central wavelength, transition probability, the energy of the lower level in the transition, the broadening coefficients, and Land\'e factors.

The observed spectra may contain only a few spectral lines, for example, SOT-SP/Hinode observations \citep[][]{HinodeSP}, or many spectral lines, such as those from TRIPPEL/SST \citep{Kiselman11}, GRIS/GREGOR \citep{Collados12}, SUSI/\textsc{Sunrise iii} \citep{Feller20,Feller25,KorpiLagg25}, SCIP/\textsc{Sunrise iii} \citep{Katsukawa26} and FISS-SP/GST \citep{vanNoort25}. \cite{Tino19} found an improvement in the reliability of inferred physical parameters when many spectral lines are used compared to only a few. However, as shown by \cite{Hoelken26}, the accuracy of the atomic data of the employed lines is extremely important to achieve this improvement. Even a few lines with poor data can lead to poor inversion results. Unfortunately, the number of spectral lines with readily available and reliable parameters is sparse compared to the number of identified spectral lines in the solar spectrum. It is important to improve the accuracy of the atomic line parameters to fully exploit the increase in the wavelength coverage of new spectrographs and deepen our understanding of the solar atmosphere.

Stars cooler than the Sun, such as K and M stars, have the peak of their radiation in the infrared. To determine the fundamental parameters of these stars, such as effective temperature, surface gravity, and the chemical composition, it is necessary to compile a line list with accurate atomic data \cite[e.g.][]{Shetrone15}. These stellar parameters are of relevance for, among other things, measuring high-precision radial velocities from stellar spectra in order to detect and precisely characterize exoplanets \citep[e.g.][]{Fisher16}, study stellar magnetic activity \citep[e.g.][]{Hahlin23}, or to study the evolution of the Milky Way center \citep[e.g.][]{Schultheis20}.

The determination of atomic line parameters has so far been achieved using one of three different approaches: laboratory measurements \citep[e.g.][]{Blackwell72,Nave94,denHartog14}, theoretical calculations from atomic structure models \citep[e.g.][]{Kurucz75,Fischer16,Bautista22}, and from observed solar \citep{GurtovenkoKostik81,GurtovenkoKostik82,Thevenin89,Thevenin90,Borrero03,Bigot06,Shchukina13,TrellesArjona21,Vukadinovic24} and stellar spectra \citep{Sigut90,Martins14,Boeche16,Laverick19}. 

Here we use spatially resolved spectropolarimetric observations of the solar atmosphere and aim to reliably infer the transition probability (denoted as \loggf{}) for many spectral lines simultaneously. We focus on the observations at 1.56\,$\mu\mathrm{m}$ obtained from the GREGOR telescope \citep{Schmidt12} using the GRIS spectrograph \citep{Collados12}. This spectral region samples the lowest layers of the solar photosphere owing to the minimum in the continuum opacity from H$^{\mathrm{-}}$ ion \citep[e.g.][]{Milic19}. The increase in the Zeeman splitting at larger wavelengths makes this spectral region a prominent choice to probe weak magnetic fields in the deepest photospheric layers \citep[see for example][]{Solanki92,Franz16,Lagg16,Borrero16,Trelles21_Hidden_Mag_Fields}. 

We apply the coupled method implemented in the \globin{}\footnote{The code is publicly available from \url{https://github.com/dvukadinovic/globin}.} inversion code \citep{Vukadinovic24} to infer the atomic parameters. The coupled method infers the atmospheric and atomic parameters self-consistently by imposing a spatial coupling in the latter, thus reducing the degeneracy between parameters. To examine the performance of the coupled method, we compare our results for \loggf{} values to those inferred in \cite{TrellesArjona21} using the same observations. This comparison is performed only for the overlapping lines, due to differences in the adopted line lists used to perform inversions.

In \sref{sec:observations} we provide a brief description of the observations. In \sref{sec:inversion} we concisely describe the coupled method and the inversion setup for the \loggf{} inference. The results are presented in \sref{sec:results} where we discuss their robustness and compare them to the results from previous studies. Additionally, in \sref{sec:mcmc} we provide a proper estimate of the uncertainties for the inferred \loggf{} values using the Markov Chain Monte Carlo method. Our conclusions are presented in \sref{sec:conclusions}.

\section{Observations}
\label{sec:observations}

We use observations from the GRIS/GREGOR spectrograph \citep{Collados12} that were obtained on the 29$^\mathrm{th}$ of August 2018 at 08:23-08:49 UTC close to the solar disc centre $(x,y)=(-185,98)^{\prime\prime}$, corresponding to the viewing angle $\theta=11.2^\circ$ ($\mu=\cos\theta=0.981$), sampling the quiet-sun region. Details of data acquisition and reduction are described in \cite{TrellesArjona21}.

\begin{figure*}[t]
    \centering
    \sidecaption
    \includegraphics[width=0.65\textwidth]{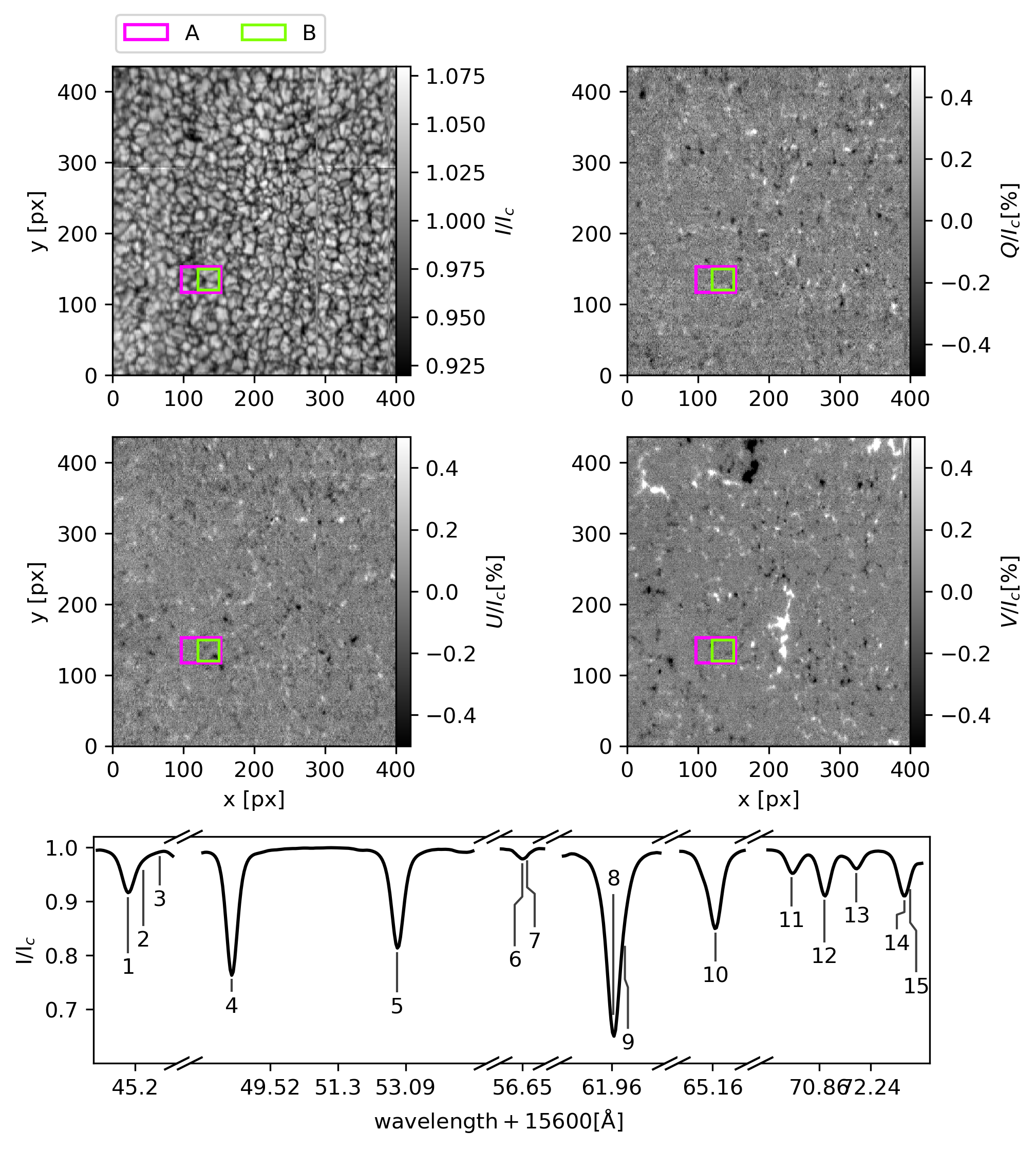}
    \caption{\textit{Top and middle panels:} Observed field of view in each Stokes component. The coloured patches represent the quiet-sun regions of interest used to infer \loggf{} values. Patch B is contained inside patch A. \textit{Bottom panel:} spatially averaged Stokes $I$ profile with line identification from \tab{tab:lines}.}
    \label{fig:fov}
\end{figure*}

\fig{fig:fov} shows the observed field of view in each Stokes component. The greyscale of the Stokes $Q$, $U$, and $V$ maps indicates the amplitude of the observed signal in fractions of the continuum intensity. The regions of interest used to infer and verify the \loggf{} values obtained in this work are marked with rectangles in different colours.

\section{Inversion setup}
\label{sec:inversion}

Height stratifications of the atmospheric physical parameters are inferred by modelling the observed profiles of spectral lines. We apply successive corrections to the atmospheric model until a match between the synthetic spectrum and the observed one is achieved. This is the core principle behind the spectropolarimetric inversion method \citep[for more details see, e.g., ][]{inversion}.

There are two different strategies for applying the inversion method to study the solar atmosphere. Commonly, the spectrum from each pixel in the field of view is analysed independently from the surrounding ones (referred to as the pixel-by-pixel inversion). Another approach, which has gained increasing importance during the last decade, is the spatially coupled inversion \citep[e.g.][]{Tiwari13,Tiwari15,Castellanos24}. The spatial coupling is imposed by limiting the variation of atmospheric parameters between neighbouring pixels \citep[spatially regularised inversion][]{Jaime19} or by using a point spread function to model light scattering in a telescope and spectrograph \citep[2D inversion][]{vanNoort12}. We impose spatial coupling via the atomic parameters, implemented in the \globin{} inversion code. We briefly describe this so-called coupled method here. A detailed description of the coupled method is presented in \cite{Vukadinovic24}.

\subsection{The coupled method}

In the coupled method, the atomic parameters for a given spectral line are restricted to be the same in all pixels within the observed field of view, which requires a simultaneous ('global') fit to all spectra. Still, the atmospheric parameters are determined solely from the fit of the spectrum from a given pixel and are therefore allowed to vary independently for each pixel. The coupled method uses the Levenberg-Marquardt method \citep{Levenberg44,Marquardt63} to find the optimal set of atmospheric and atomic parameters that match the observed spectra through the minimization of the $\chi^2$ merit function:
\begin{equation}
    \chi^2(\mathbf{p}) = \frac{1}{N-N_\mathrm{p}}\sum_{a=1}^{N_\mathrm{atm}}\sum_{i=1}^N \frac{w_{i}^2}{\sigma_i^2} \cdot \left( \mathbf{O}_{i,a} - \mathbf{I}_{i,a}(\mathbf{p})  \right)^2,
    \label{eq:chi2}
\end{equation}
where $\mathbf{O}$ represents the observed Stokes spectra and $\mathbf{I}$ the synthetic Stokes spectra. The Levenberg-Marquardt method allows for adjustment of weights, $w$, and the noise level, $\sigma$, for each Stokes component and wavelength. The index $a$ runs over every pixel in the observed field of view, whereas index $i$ runs over each wavelength and Stokes component ($N=4\cdot N_\mathrm{w}$). The vector $\mathbf{p}$ contains the atmospheric parameters for every individual pixel in the field of view as well as the atomic parameters $\mathbf{p} = (\mathbf{p}_1, \mathbf{p}_2, \ldots, \mathbf{p}_{N_\mathrm{atm}},\mathbf{p}_\mathrm{global})$, where $N_\mathrm{atm}$ is the total number of pixels in the field of view. The total number of free parameters used to minimize the $\chi^2$ merit function is $N_\mathrm{p}$.

The minimisation of the $\chi^2$ with the Levenberg-Marquardt method is achieved through an iterative solution of:
\begin{equation}
    \mathcal{H_\mathrm{global}}\cdot\Delta \mathbf{p_j} = \mathcal{J}_\mathrm{global}^\mathrm{T}\cdot\mathbf{\Delta},
    \label{eq:globin_LM}
\end{equation}
from which we obtain the correction of the free parameters in the $j$-th step, $\Delta \mathbf{p_j}$. Matrix $\mathcal{J}_\mathrm{global}$ is the Jacobian matrix constructed from the response functions\footnote{Response function is the first derivative of the Stokes spectrum with respect to a physical parameter.} of the Stokes spectra for each free parameter. It has a semi-block-diagonal form originating from the response functions to the atmospheric parameters for each pixel in the field of view. The last $n_\mathrm{global}$ columns of the Jacobian matrix contain the response functions to $n_\mathrm{global}$ parameters for which spatial coupling is imposed, such as the atomic parameters. The Hessian matrix is constructed from the Jacobian as $\mathcal{H} = \mathcal{J}^\mathrm{T}\mathcal{J} + \lambda_\mathrm{M}\cdot \mathrm{diag}(\mathcal{J}^\mathrm{T}\mathcal{J})$, where $\lambda_\mathrm{M}$ is the Marquardt parameter that controls how fast we converge to the $\chi^2$ minimum. The term $\mathbf{\Delta}$ in \eq{eq:globin_LM} represents the differences between the observed and synthetic Stokes spectra.

\subsection{Inversion strategy}

Modelling the synthetic spectrum for comparison with the observed one requires a finely stratified atmospheric model. From a practical point of view, inferring the physical parameters of such a finely stratified atmosphere is numerically demanding and leads to a less constrained inversion \citep{Jaime19_STiC}, partly due to degeneracies among the inferred parameters \citep[see, e.g.,][]{inversion}. Thus, the usual approach is to simplify the atmospheric model and infer its physical parameters only at particular heights in the atmosphere called nodes.

The \globin{} code parametrizes the atmospheric structure using nodes that represent the values of different physical parameters at specific atmospheric heights. The number of nodes and their positions are specified for each individual parameter. The nodes are distributed over height points at which the observed lines are formed. By default, the node positions are assumed to be given in the optical depth scale at $500$\,nm. The finer stratification of physical parameters is then obtained by interpolation with cubic Bezier polynomials, assuming an optical depth resolution of $\Delta\log\tau_{500}=0.1$.

From the finely interpolated temperature stratification and assuming an atmospheric model in hydrostatic equilibrium, we calculate the electron number density and gas pressure assuming the ideal gas equation and solar chemical composition from \cite{Asplund21}. The emergent spectrum from the interpolated atmospheric model is calculated with a wrapper around the \texttt{RH} code built into the \globin{} code.

We chose to infer temperature, line of sight velocity, micro-turbulent velocity, and magnetic field vector (strength, inclination, and azimuth), whose node positions are listed in \tab{tab:nodes}. We have omitted the temperature node at $\log\tau_{500}=0$ since this wavelength region is formed deeper in the atmosphere. Thus, we opted for the temperature node at $\log\tau_{500}=0.2$ for the base of the photosphere.

Inference of the \loggf{} values from the entire observed field of view creates a large parameter space, which impacts the minimization of the $\chi^2$ merit function to find the optimal set of free parameters that describe the observed spectra. Assuming that the field of view has dimensions of $400\times400$ and that for each pixel we have $10$ atmospheric parameters (a conservative estimate), we end up with $1.4\cdot10^6$ parameters. We are thus required to invert a Hessian matrix of $1.4\cdot10^6 \times1.4\cdot10^6$ elements in each inversion step to obtain the parameter correction $\Delta \mathbf{p_j}$ from \eq{eq:globin_LM}. Inverting such a matrix is computationally demanding and prone to significant numerical artefacts. This makes the method inapplicable to larger fields of view, such as the one from the GRIS/GREGOR spectrograph. We have to select smaller regions of interest to which we can apply the coupled method and guarantee the convergence of the parameters. The regions of interest used in this work to infer the \loggf{} values are shown in \fig{fig:fov}. The patches A and B sample the quiet-sun region, where patch B is completely contained inside patch A.

\begin{table}[]
    \centering
    \caption{Node positions of atmospheric parameters in $\log\tau_\mathrm{500}$.}
    \begin{tabular}{c c}
        \toprule[1.5pt]
        parameter & nodes \\ \midrule[1.5pt]
        T &  -1.5, -0.9, -0.4, 0.2 \\
        $\varv_\mathrm{LOS}^{m}$, $v_\mathrm{mic}^{m}$& constant \\
        B & -1.0, 0 \\
        $\theta$, $\phi$ & constant \\
        $\varv_\mathrm{LOS}^{nm}$, $v_\mathrm{mic}^{nm}$, $\alpha$ & constant \\
        \bottomrule[1.5pt]
    \end{tabular}
    \label{tab:nodes}
\end{table}

The observed spectra from ground-based observatories suffer from scattered light on their way through the Earth's atmosphere (atmospheric seeing) and inside the telescope and spectrograph. These scatterings affect the line shapes and, if not treated properly, lead to poor inference of free parameters \citep{Saranathan21}. There are different ways to mitigate them: either removing them in the data reduction process or modelling them during the inference process. Here we chose the second approach and modelled them using a two-component atmospheric model. In this case, the synthetic spectrum that is used to fit the observations is a combination of radiation from the magnetised atmospheric model, $\mathbf{I}^\mathrm{m}$, and the unpolarized scattered light emitted from the non-magnetised atmosphere, $\mathbf{I}^\mathrm{nm}$:
\begin{equation}
    \mathbf{I} = (1-\alpha)\mathbf{I}^\mathrm{nm} + \alpha\mathbf{I}^\mathrm{m},
\end{equation}
where $\alpha$ is a magnetic filling factor, which was mainly introduced to counter the effects of stray light. We opted to use the Harvard-Smithsonian reference atmospheric model \citep[HSRA;][]{HSRA} without chromospheric temperature rise as the non-magnetised atmospheric model with a free depth-independent line of sight, $v_\mathrm{LOS}^{nm}$, and micro-turbulent, $v_\mathrm{mic}^{nm}$, velocities. The synthetic spectrum is additionally convolved with a Gaussian function with the standard deviation corresponding to $v_\mathrm{mac}=1.56$\,km/s to mimic (mainly) the instrumental broadening as well as macro-turbulent broadening.

The inference of \loggf{} values is achieved by using at least one spectral line with a well-determined \loggf{} value that is strong and sensitive enough to constrain the atmospheric structure \citep[reference line;][]{Vukadinovic24}. In the spectral range considered here, line 15648\,\AA{} has an experimentally measured value \citep{Ruffoni13}, qualifying it to be used as the reference line. For all other spectral lines, \loggf{} is varied to find the best fit to the observed spectra.

Another important parameter that typically should not change across the field-of-view is the chemical abundance of the element from which the spectral line originates. Abundance and \loggf{} enter the computation of the spectral profile as a common factor and therefore alter the strength of a spectral line in a similar way. The determination of \loggf{} and abundance from a single atomic species spectrum is a degenerate problem. This degeneracy can only be lifted by fixing either one of the two to previous measurements, or to some extent by analysing spectra containing lines from many different atomic species. In the spectral region used for this work, we mainly find Fe\,I lines, two Si\,I lines and one S\,I line. The elemental abundances are taken from \cite{Asplund21}, which are $A_\mathrm{Fe}=7.46$, $A_\mathrm{Si}=7.51$, and $A_\mathrm{S}=7.12$, and are assumed to be constant throughout the inversions. Except for the atomic lines, we also had to include $A-X$ $^{12}$C$^{14}$N lines from \cite{Kozlov24}, which show a non-negligible opacity contribution.

\tab{tab:lines} lists the important atomic parameters used to model the observed spectral lines under the assumption of local thermodynamic equilibrium because of their high excitation potential. These lines thus form in deep photospheric layers where collisions between particles are frequent, and the radiation field does not impact level populations. In the same table, we also list the \loggf{} values reported by other authors, corrected by the differences in the elemental abundance assumed by those authors relative to the values used here. The abundance correction allows a direct comparison between inferred \loggf{} values derived assuming different chemical compositions of the solar atmosphere.

The van der Waals broadening of spectral lines is computed using the \texttt{abo-cross} software \citep{ABO98_code,ABO_code,ABO95,ABO97,ABO98}.

\section{Inversion results}
\label{sec:results}

All inversions of different regions of interest were performed with the same initial values and convergence criteria. To attain convergence, each inversion was run in four cycles. The first three cycles were given a maximum of 10 iterations, while the last one had 30 iterations. All inversion runs converged in the last cycle, indicated by a small change in the $\chi^2$ value in the last iteration step. Between the cycles, the atmospheric parameters of both components were spatially smoothed with a Gaussian function. The smoothing process alters the initial conditions for the next cycle in a way that prevents the optimizer from getting stuck in a local minimum in the $\chi^2$ hypersurface. This smoothing process does not apply to the global \loggf{} values, since they are the same for all spatial pixels. The optimum strategy for the global parameters was to use the weighted average of values from the previous cycle as a starting solution for the following cycle. The values from the last iteration have a higher weight than the initial values.

The weighting of the Stokes components in \eq{eq:chi2} was $w_I=1$, $w_Q=1.5$, $w_U=1.5$, and $w_V=1.5$, consistent for all inversions in the present study. We estimated the noise from the observed continuum to be $\sigma=10^{-3}$ and used it to compute the $\chi^2$ from \eq{eq:chi2}.

\subsection{Reference inversion}
\label{sec:inv_patch_A}

\begin{figure}[ht!]
    \centering
    \includegraphics[width=0.8\linewidth]{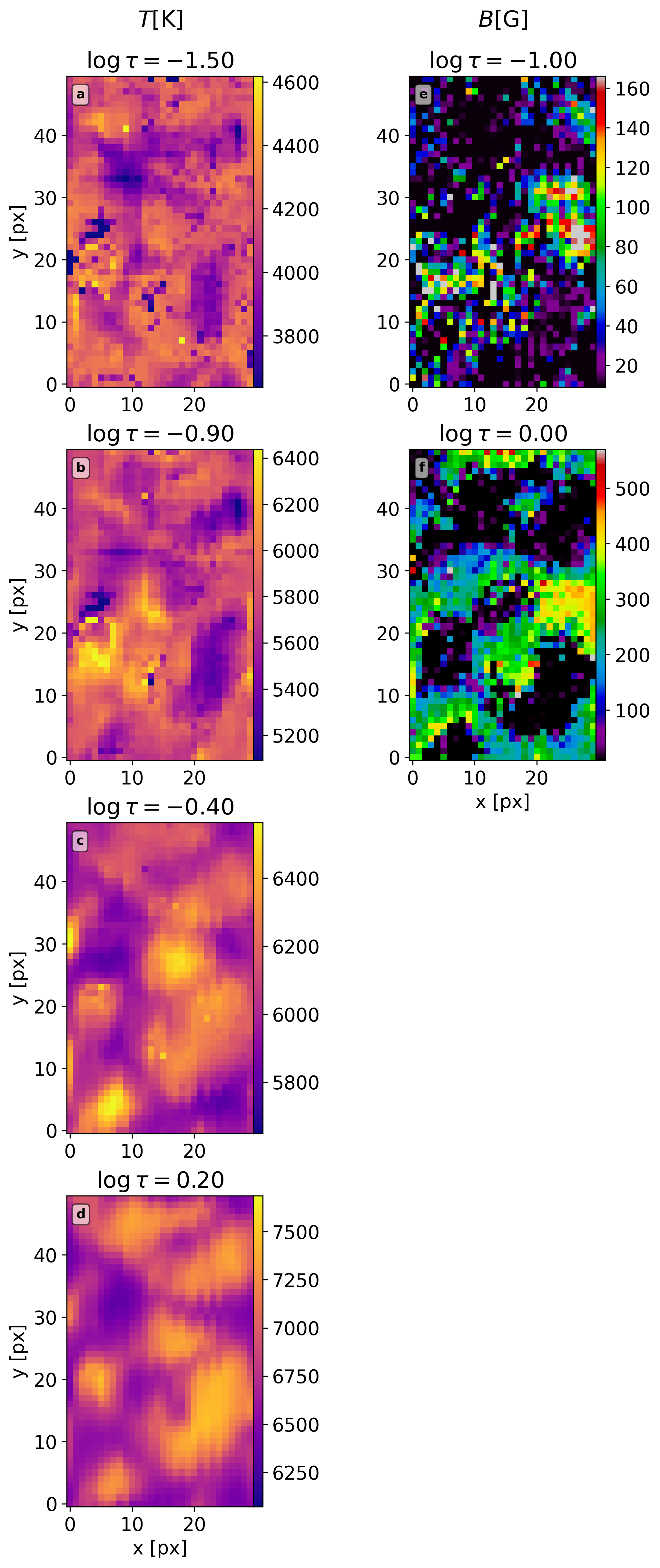}
    \caption{Inferred depth-dependent atmospheric parameters for patch A: temperature and magnetic field strength. Node positions are listed at the top of each panel. The x- and y-axes indicate pixel positions in the patch.}
    \label{fig:atmos_depth_dependent}
\end{figure}

\begin{figure}[ht!]
    \centering
    \includegraphics[width=0.8\linewidth]{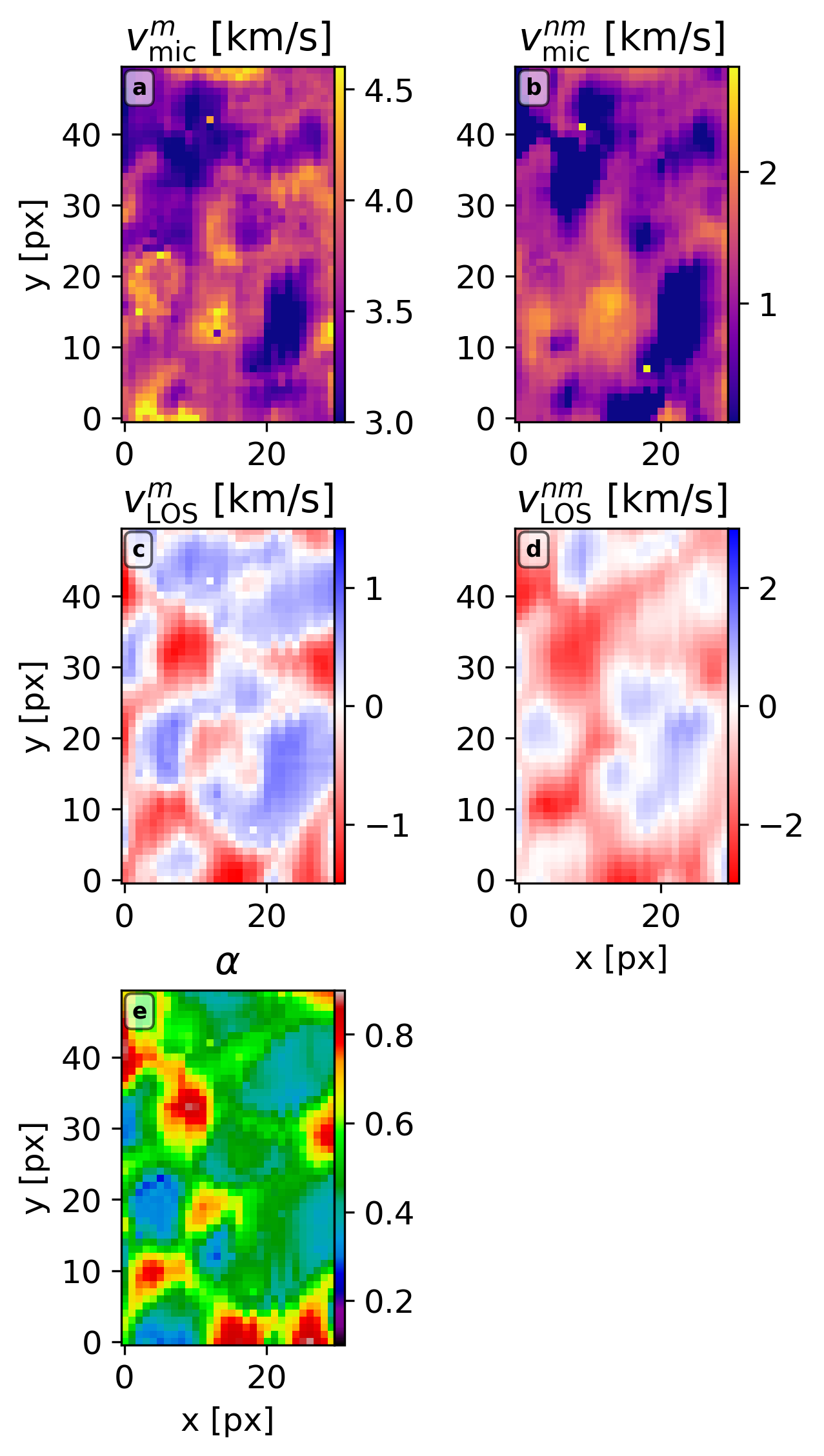}
    \caption{Inferred depth-independent atmospheric parameters for patch A: micro-turbulent velocity of the magnetised component, micro-turbulent velocity of the non-magnetised component, LOS velocity of the magnetised component, LOS velocity of the non-magnetised component, and magnetic filling factor. The x- and y-axes indicate pixel positions in the patch.}
    \label{fig:atmos_depth_independent}
\end{figure}

Here we present inversion results from patch A that defines our reference case to draw conclusions on the strength of the coupled method to determine the \loggf{} values. The inferred atmospheric parameters of both atmospheric components are displayed in \fig{fig:atmos_depth_dependent} for the depth-dependent and in \fig{fig:atmos_depth_independent} for the depth-independent parameters. We do not show the magnetic field inclination and azimuth because, in most of the pixels, there was insufficient signal in the observed spectra for a reliable determination. The inferred \loggf{} values are listed in Table\,\ref{tab:loggf} and will be discussed in more detail in the next section. Here, we discuss the inference of atmospheric parameters.

\begin{table}[]
    \centering
    \caption{Inferred \loggf{} values from the reference inversion.
    }
    \begin{tabular}{c l c c}
    \midrule[1.5pt]
    No. & Element & $\lambda$ [\AA{}] & \loggf{} \\ \midrule[1.5pt]
    1 & Fe I & 15645.016 & -0.368 $\pm$ 0.004 \\ 
	2 & Si I & 15645.415 & -1.281 $\pm$ 0.015 \\ 
	3 & Si I & 15645.843 & ... \\ \midrule[1pt]
	4 & Fe I & 15648.510 & -0.61\tablefootmark{1} \\ \midrule[1pt]
	5 & Fe I & 15652.871 & 0.013 $\pm$ 0.003 \\ \midrule[1pt]
	6 & Fe I & 15656.632 & -1.517 $\pm$ 0.011 \\ 
	7 & Fe I & 15656.767 & ... \\ \midrule[1pt]
	8 & Fe I & 15662.013 & 0.176 $\pm$ 0.003 \\ 
	9 & Fe I & 15662.313 & -0.786 $\pm$ 0.014 \\ \midrule[1pt]
	10 & Fe I & 15665.240 & -0.374 $\pm$ 0.002 \\ \midrule[1pt]
	11 & Fe I & 15670.124 & -0.789 $\pm$ 0.005 \\ \midrule[1pt]
	12 & Fe I & 15671.004 & -0.348 $\pm$ 0.003 \\ \midrule[1pt]
	13 & Fe I & 15671.866 & -1.169 $\pm$ 0.006 \\ \midrule[1pt]
	14 & Fe I & 15673.151 & -0.450 $\pm$ 0.005 \\
	15 & S I & 15673.308 & -1.185 $\pm$ 0.048 \\ 
    \midrule[1.5pt]
    \end{tabular}
    \tablefoot{The spectral lines are sorted by ascending wavelength. The columns represent the line number in the analysed spectral region, the chemical element and ionisation stage, the central wavelength of the line, and \loggf{} values with uncertainty determined in the present work. Elemental abundances are taken from \cite{Asplund21}, and are $A_\mathrm{Fe}=7.46$, $A_\mathrm{Si}=7.51$, and $A_\mathrm{S}=7.12$. Different blend pairs are separated by horizontal lines.\\
    References. \tablefoottext{1}{\cite{Ruffoni13}}
    }
    \label{tab:loggf}
\end{table}

The temperature maps of the magnetised component in the top two nodes show some salt-and-pepper patterns and discontinuities (see, e.g., the purple patch at $(x,y)=(6,25)$ in \fig{fig:atmos_depth_dependent}a), which typically run along the edges of magnetic features. These artefacts are a result of increased parameter degeneracy in pixels with no signal above the noise level in the Stokes $Q$, $U$, or $V$ profiles. This is the case for most of the bright pixels inside the granules. There, the inferred magnetic field strength is low (see \fig{fig:atmos_depth_dependent}e-f panels), thus removing constraints on the magnetic filling factor. The inversion code preferred this scenario and lower values for the magnetic filling factor inside granules (see \fig{fig:atmos_depth_independent}e). The inferred magnetic filling factor is on average 52\,\%, reaching values up to 82\,\%. As soon as a measurable polarization profile is detected, this degeneracy is removed, and the inversion code differentiates between the magnetic field and the magnetic filling factor contributions, thus improving the inference of the temperature stratification.

The magnetic field in the quiet-sun regions is concentrated in the intergranular lanes, with a field strength of a few hundred Gauss (\fig{fig:atmos_depth_dependent}f). These pixels are reproduced only by adding the non-magnetised atmospheric component. Without the non-magnetic component, the field strength is unable to simultaneously reproduce the Zeeman broadening in unpolarized line profiles and the polarization recorded in Stokes $Q$, $U$, and $V$ profiles.

The micro-turbulent velocities of both atmospheric components mutually exchange their values, preserving the line width (panels a and b in \fig{fig:atmos_depth_independent}). These pixels also coincide very well with those from the temperature panels, suffering from the increased degeneracy between inferred parameters as the polarization signal is buried in the noise.

\begin{figure*}[h!]
    \centering
    \sidecaption
    \includegraphics[width=0.65\linewidth]{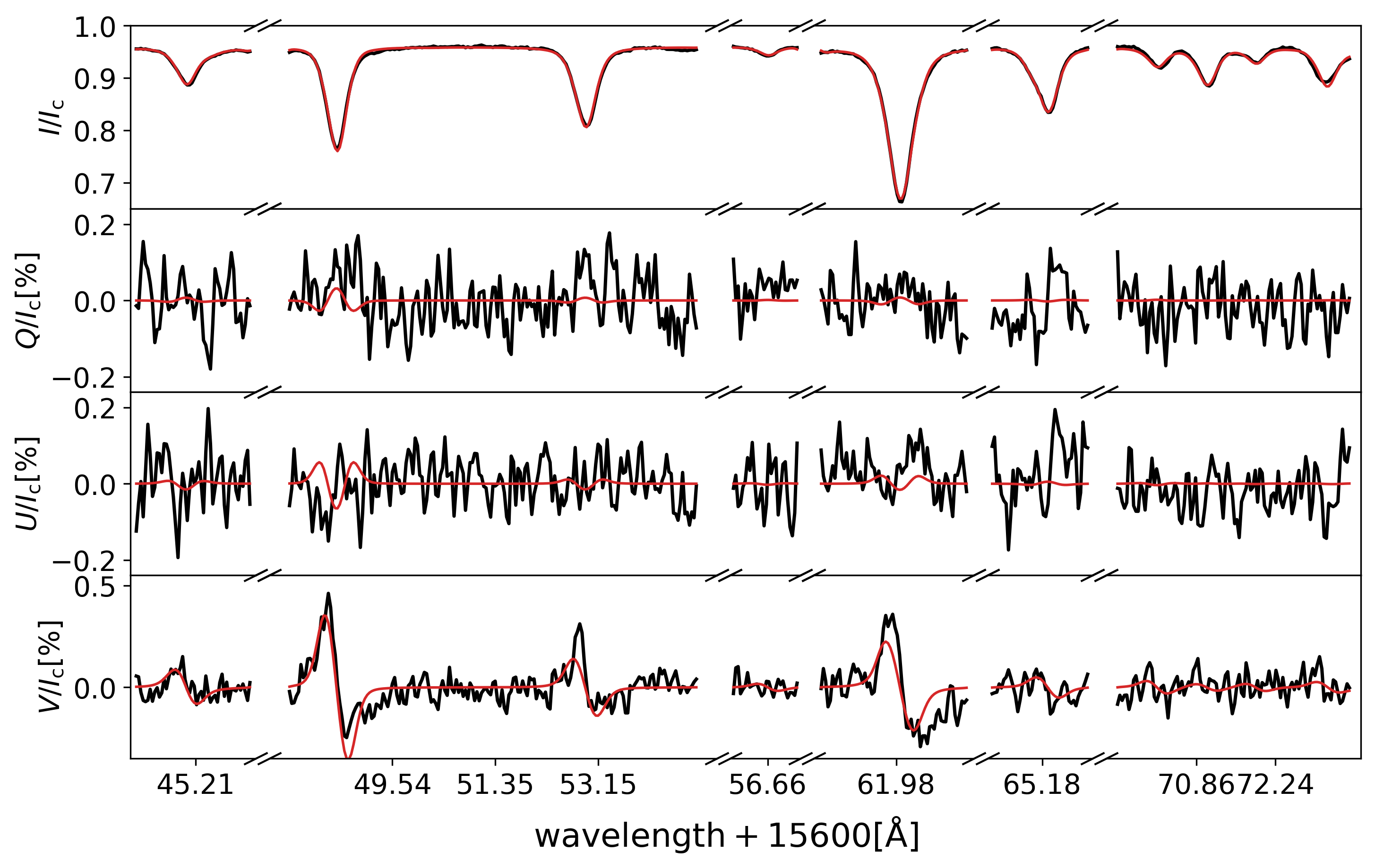}
    \caption{The achieved fit (red line) to an observed spectrum (black line) from a pixel with $\chi^2=3.08$ (the median value of $\chi^2$ across patch A).}
    \label{fig:stokes_fit}
\end{figure*}

The line of sight velocity maps for the magnetised and non-magnetised components show a characteristic granular pattern observed in the photosphere (panels c and d in \fig{fig:atmos_depth_independent}). Since we opted for height-independent inference of the line of sight velocities in both atmospheric components, the inversion code is limited in reproducing the observed line asymmetry. Combining the spectra from magnetized and non-magnetized components partially reproduces the Stokes $I$ line asymmetry due to different temperature stratifications. However, the line asymmetry in polarized profiles is poorly reproduced since we did not allow for the velocity gradients in the atmosphere (see \fig{fig:stokes_fit}). Additionally, the lack of polarization signals hinders the inference of a more complex line of sight stratification. We performed a test with three nodes for the line of sight velocity of the magnetized component. We observed improvements of the fit quality in some of the pixels, but the overall fit quality has not improved. Thus, we decided to retain a single node for the line of sight velocity.

\begin{figure}[h!]
    \centering
    \includegraphics[width=0.8\linewidth]{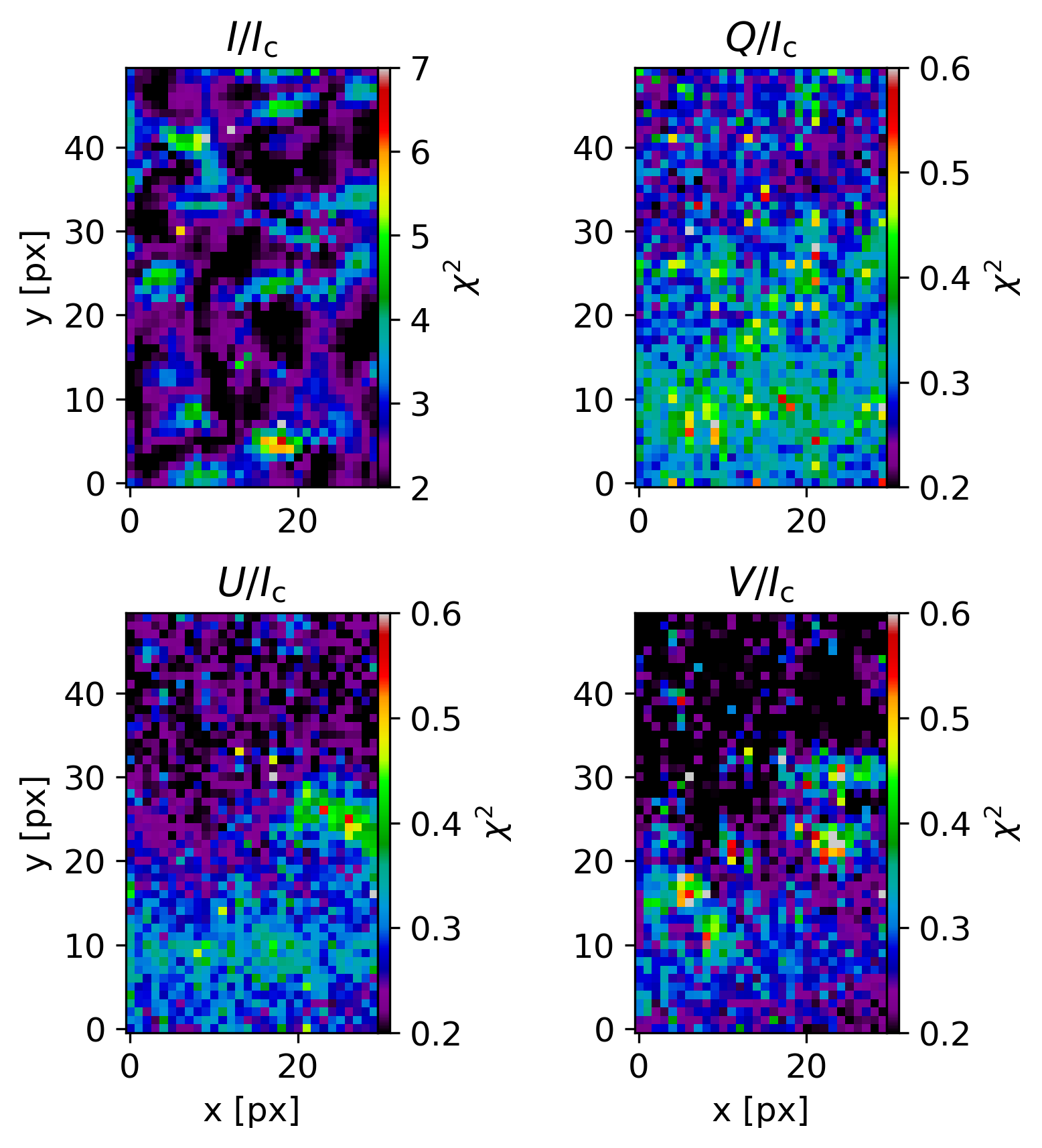}
    \caption{RMS differences between the observed and the best-fit spectrum for each Stokes component. The fit of the Stokes $I$ component dominates over the polarized components (as expected from the inversion of quiet-sun profiles). The x- and y-axes indicate pixel positions in the patch A.}
    \label{fig:chi2_A}
\end{figure}

The root-mean-square (RMS) differences between the observed and synthesized Stokes profiles weighted by the continuum intensity are displayed in \fig{fig:chi2_A} for each Stokes component. The fit of individual pixels, allowing for different \loggf{} values in each pixel,  increases flexibility to lower the $\chi^2$ further. However, this comes with the potential for erroneous inference of atmospheric parameters, as the atomic ones would be very different for each pixel. The spatially coupled inversions have to deal with a large parameter space. This increases the number of local minima in the $\chi^2$ that are of very similar depth, but with the advantage of ensuring the reliable inference of the atomic parameters. These atomic parameters can, in a second step, be used to retrieve the atmospheric parameters using other inversion methods, such as the spatially regularised or 2D inversion. 

We display the fitted and observed spectrum in a pixel with a typical $\chi^2$ value in \fig{fig:stokes_fit}. Stokes $Q$ and $U$ are throughout the considered patch buried in the noise; only the Stokes $V$ profiles are strong enough to stand above the noise and are reproduced to some extent. The lack of asymmetries in the inferred Stokes $V$ profiles is because of the height-independent line of sight velocity inference. On the other hand, the Stokes $I$ profiles and the continuum are well reproduced. Small deviations from the fitted profile in the core of some lines are observed in some pixels. This is, however, not a result of poorly inferred \loggf{} values for these lines, but rather a consequence of a more complex atmospheric stratification in these pixels, which cannot be reproduced with the simple model atmosphere used here. The treatment of the \loggf{} values as global parameters makes their determination robust, even when some pixels cannot be fit perfectly.

\subsection{Robustness of inferred \loggf{} values against inversion setup and observational effects}

In this section, we discuss five cases that test the robustness of the \loggf{} inference and contrast the results with the reference case presented in the previous section.

\begin{itemize}[itemsep=0.5em]

\item[] \textbf{Case I.} The effect of the selected region of interest. We chose to compare the results from two patches, A and B, with B lying inside A (see Fig.\,\ref{fig:fov}). Atmospheric parameters are compared only for the overlapping pixels.

\item[] \textbf{Case II.} The effect of not correcting for spectral stray light. The spectral stray light correction alters the relative line depth, which impacts the inference of \loggf{} values. We subtracted 3.5\% of the spectral stray light from the observed spectra in patch A, for which we repeated the inversion.

\item[] \textbf{Case III.} The effect of choosing a different reference line. We repeated the inversion of patch A using the 15662\,\AA{} line as the reference one. To properly contrast the results with the reference case, we set the \loggf{} value for the 15662\,\AA{} line to its inferred value from the reference case. This case aims to demonstrate that the coupled method is able to retrieve comparable atmospheric structures and \loggf{} values irrespective of which line with a reliably determined \loggf{} value we opt for. 

\item[] \textbf{Case IV.} The effect of the selected atmospheric parametrisation. The choice of the number of nodes for each parameter and their positions affects an atmospheric model's ability to reproduce observed complex line profiles and influences the inferred \loggf{} value. We opted to test how strong such an influence is by decreasing the number of nodes for the temperature stratification to three and placing them at $\log\tau_\mathrm{500}=(-1.5, -0.6, 0.2)$. We decided to keep the same top and bottom nodes while placing the middle node in between.

\item[] \textbf{Case V.} The effect of missing lines. Disregarding blended lines could interfere with the inferred atomic parameters of their blend pair. We have repeated the inversion of patch A after removing lines 2, 3, 7, 9, and 15 from the line list.

\end{itemize}

We computed the mean and standard deviation of the difference between parameters inferred in each case and those inferred from the reference case. The values for \loggf{}, temperature, line of sight and macro-turbulent velocity in the magnetized component, magnetic filling factor, and magnetic field strength are presented in the \tab{tab:std_values}. The standard deviation for the height-dependent parameters was computed using the interpolated values over the atmospheric layers spanned by nodes (extrapolated values are excluded from the comparison). 

\begin{table*}[h!]
    \centering
    \caption{The mean and standard deviation of the differences between the parameter values in a given case and those from the reference inversion.}
    \begin{tabular}{l d{-1} d{-1} d{-1} d{-1} d{-1} d{-1}}
        \midrule[1.5pt]
        cases & \multicolumn{1}{c}{$\log(gf)$} & \multicolumn{1}{c}{$T$ [K]} & \multicolumn{1}{c}{$v^m_\mathrm{LOS}$ [km/s]} & \multicolumn{1}{c}{$v_\mathrm{mic}^m$ [km/s]} & \multicolumn{1}{c}{$\alpha$ [\%]} & \multicolumn{1}{c}{$B$ [G]} \\
        \midrule[1.5pt]
        Case I & -0.013 ? 0.017 & 81 ? 240 & -0.04 ? 0.05 & 0.11 ? 0.22 & -1 ? 3 & 46 ? 243 \\
        Case II & 0.010 ? 0.014 & -206 ? 271 & -0.02 ? 0.09 & -0.15 ? 0.20 & 5 ? 4 & -201 ? 269 \\
        Case III & -0.006 ? 0.037 & 102 ? 539 & -0.13 ? 0.08 & -0.4 ? 0.31 & 4 ? 7 & -3 ? 607 \\
        Case IV & 0.005 ? 0.012 & -70 ? 289 & 0.04 ? 0.07 & 0.42 ? 0.30 & -4 ? 4 & 23 ? 260 \\
        Case V & -0.002 ? 0.025 & 164 ? 384 & -0.64 ? 0.20 & 1.07 ? 0.43 & -9 ? 4 & 142 ? 402 \\
        \midrule[1.5pt]
    \end{tabular}
    \tablefoot{The columns are for the \loggf{}, temperature, line of sight and micro-turbulent velocity of the magnetized component, magnetic filling factor, and magnetic field strength.}
    \label{tab:std_values}
\end{table*}

\begin{figure}[h!]
    \centering
    \includegraphics[width=0.85\linewidth]{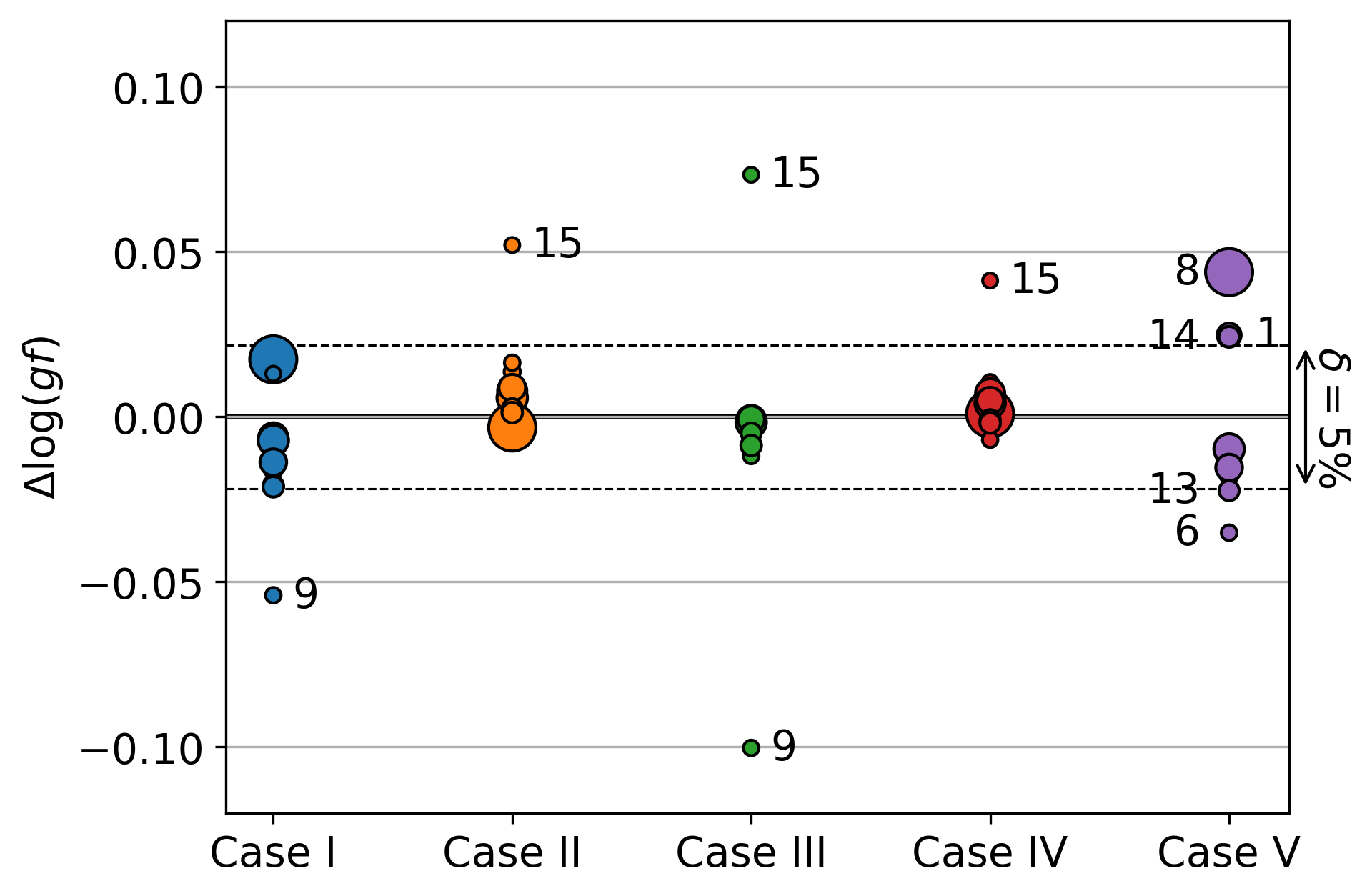}
    \caption{Differences in inferred \loggf{} values in comparison to the reference inversion. Stronger lines are represented with larger circles. The lines showing \dloggf{} larger than 5\,\% (indicated by the horizontal dashed lines) are marked with their respective line number.}
    \label{fig:loggf_vs_cases}
\end{figure}

In \fig{fig:loggf_vs_cases}, we show the differences in the inferred \loggf{} values for each case considered here. In all cases except case V, the inferred \loggf{} values agree well with those from the reference inversion, with deviations of less than 5\%, particularly for weak blended lines, despite variations in the atmospheric parameters. This demonstrates the robustness of the coupled method for inferring \loggf{} values, even in the presence of inaccuracies in the data reduction process or differences in the chosen inversion setup. This conclusion holds provided that a basic data reduction is performed and the inversion method is applied as intended.

Case I has one of the smallest deviations in both the inferred atmospheric parameters and the \loggf{} values, compared to the reference case. This shows that the coupled method is insensitive to the selection of the field of view. The advantage of the coupled method is that pixels need not be spatially coherent. This allows flexibility to analyse distant pixels that represent a variety of solar features observed in a large field of view. 

The correction of the spectral stray light is applied to each independent pixel by the same amount (case II). This correction has a minor impact on the inferred atomic parameters, but significant deviations are observed in temperature, magnetic filling factor, and magnetic field strength. The small deviations in the inferred atomic parameters are a consequence of the coupled method, which infers the relative \loggf{} values correctly. The relative \loggf{} is the difference between the inferred \loggf{} and the \loggf{} value of the reference line. However, the coupled method compensates for the different spectral stray light compared to the reference inversion by adjusting the atmospheric parameters. As long as we have pixel-dependent effects/corrections that alter the spectral line's depth, the atomic parameters will still be recovered correctly, whereas the atmospheric parameters would be altered to compensate for the applied correction.

Changing the reference line (case III) alters the response of the Stokes profiles to the atmospheric parameters and affects their inference. The inferred atmospheric and atomic values do not show large deviations from the reference case. The only exception is a larger variation in the temperature and magnetic field strength (and thus of the magnetic filling factor), which could come from a different temperature and magnetic sensitivity of lines 15648\,\AA{} and 15662\,\AA{}. This example further supports the premise that the coupled method preserves the relative \loggf{} values. As long as the chosen reference line has a reliably determined \loggf{} value (e.g., from accurate experimental measurements), we will retrieve comparable atmospheric parameters. Here we have a case with only one line that satisfies this criterion. For a broader spectral region, typically, more lines could meet this criterion, which would further assist in disentangling the atmospheric and atomic parameters. 

Altering the temperature stratification (case IV) of the magnetised atmosphere produced the smallest deviation and variation in the inferred atmospheric parameters. Temperature is the atmospheric parameter that acts most similarly to \loggf{} by changing line strength, in particular, of weaker lines of neutral species, so that \loggf{} has the strongest degeneracy with temperature. However, in this case, the three-node atmospheric model can provide an equally good fit to the observed spectra as the four-node model. We note an increase in the micro-turbulent velocity of the magnetized component in this case, which is tightly correlated with the temperature. 

Great effort is directed toward identifying spectral lines and assigning them to the correct atomic species and obtaining their fundamental parameters \citep[e.g.,][to mention a few]{Kurucz75, Nave94, Castelli10, Li26}. Disregarding lines, especially weak ones, can cause a measurable change in the inferred parameters. For example, if the weak line lies in the wing of a stronger line, ignoring it must be compensated by increasing the \loggf{} value of the strong line, or by adjusting atmospheric parameters.

In case V, we note significant changes in \loggf{} values of strong lines (1, 8 and 14) whose blends were disregarded here. Due to the lack of blend components, their \loggf{} values are increased to compensate for the missing opacity in their wings. Furthermore, because of the \loggf{} adjustments, the atmospheric parameters need to be altered as well due to spatial coupling of the atomic parameters. Here, we observe a significant increase in the micro-turbulent velocity of the magnetized component that aims to broaden lines and compensate for the lack of opacity in the wings of strong lines. Additionally, the line of sight velocity is decreased, which causes a net redshift of the lines since all omitted ones are in the red wings of the stronger blend pair. Simultaneously, there is a large decrease in the magnetic filling factor. However, as these changes in the atmospheric parameters cause an increase in the opacity, the \loggf{} of other lines is decreased to compensate for it. Disregarding a weak blend line can have an indirect effect on the \loggf{} of a line that is not part of its blend by affecting the atmospheric parameter inference with the coupled method. 

\begin{figure*}[h!]
    \centering
    \includegraphics[width=0.8\textwidth]{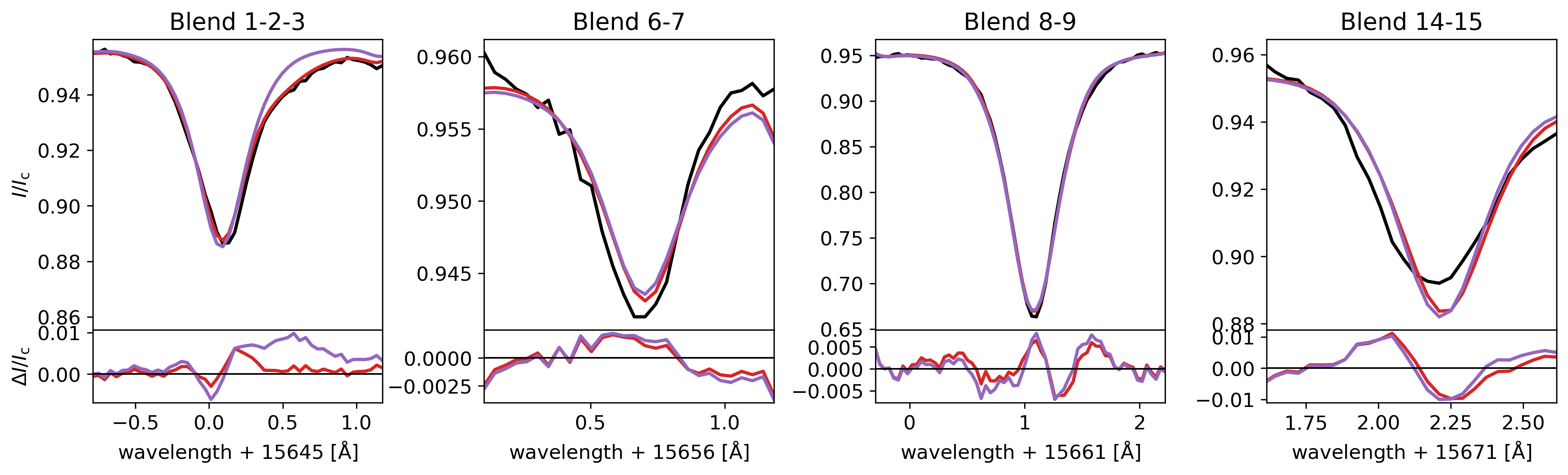}
    \caption{Spectral regions around blended pairs used in the case of the missing blended lines. We show only the Stokes $I$ fit, as these lines have polarization signal below the noise level. The black line shows the observed spectrum, the red line is the best-fit spectrum obtained in the reference inversion, and the purple line shows the best-fit spectrum from case V, where we excluded lines 2, 3, 7, 9, and 15 from the line list. The spectrum is shown for the same pixel as in \fig{fig:stokes_fit}.}
    \label{fig:missing_lines_fit}
\end{figure*}

Removing these weak lines from the line list caused an increase of $26\%$ in the $\chi^2$. The effect of excluding line 2 from the line list is the largest owing to its strength and broad wings caused by the Stark broadening (\fig{fig:missing_lines_fit}). Excluding lines 7, 9, and 15 shows almost no difference to the reference inversion, indicating poor constraints on their \loggf{} value from the inversion of the observation set we use. This is already seen in Fig.\,\ref{fig:loggf_vs_cases} where lines 9 and 15 show the largest deviations from the reference case.

Lines 3 and 7 were excluded from any previous discussion because their deviation exceeds 1 dex. These large differences arise because the lines are too weak to produce a significant change in $\chi^2$ and therefore have little impact on the fit quality. This result is further supported by the Markov chain Monte Carlo sampling (see \sref{sec:mcmc}) where the joint posterior distribution for their \loggf{} is flat. For this reason, we decided to omit the inferred values of these lines from \tab{tab:lines}.

We also want to highlight here that case V does not necessarily check the completeness of a line list. An inversion code aims to reproduce observations the best it can using the provided line list, chosen atmospheric parameterization and implemented physical constraints. It is up to users then to properly bound free parameters and verify that inferred values reflect the reality. In the next section, we briefly discuss the completeness of the adopted line list.

\subsection{Comparison with previous studies}
\label{sec:comparison}

\begin{figure}
    \centering
    \includegraphics[width=0.85\linewidth]{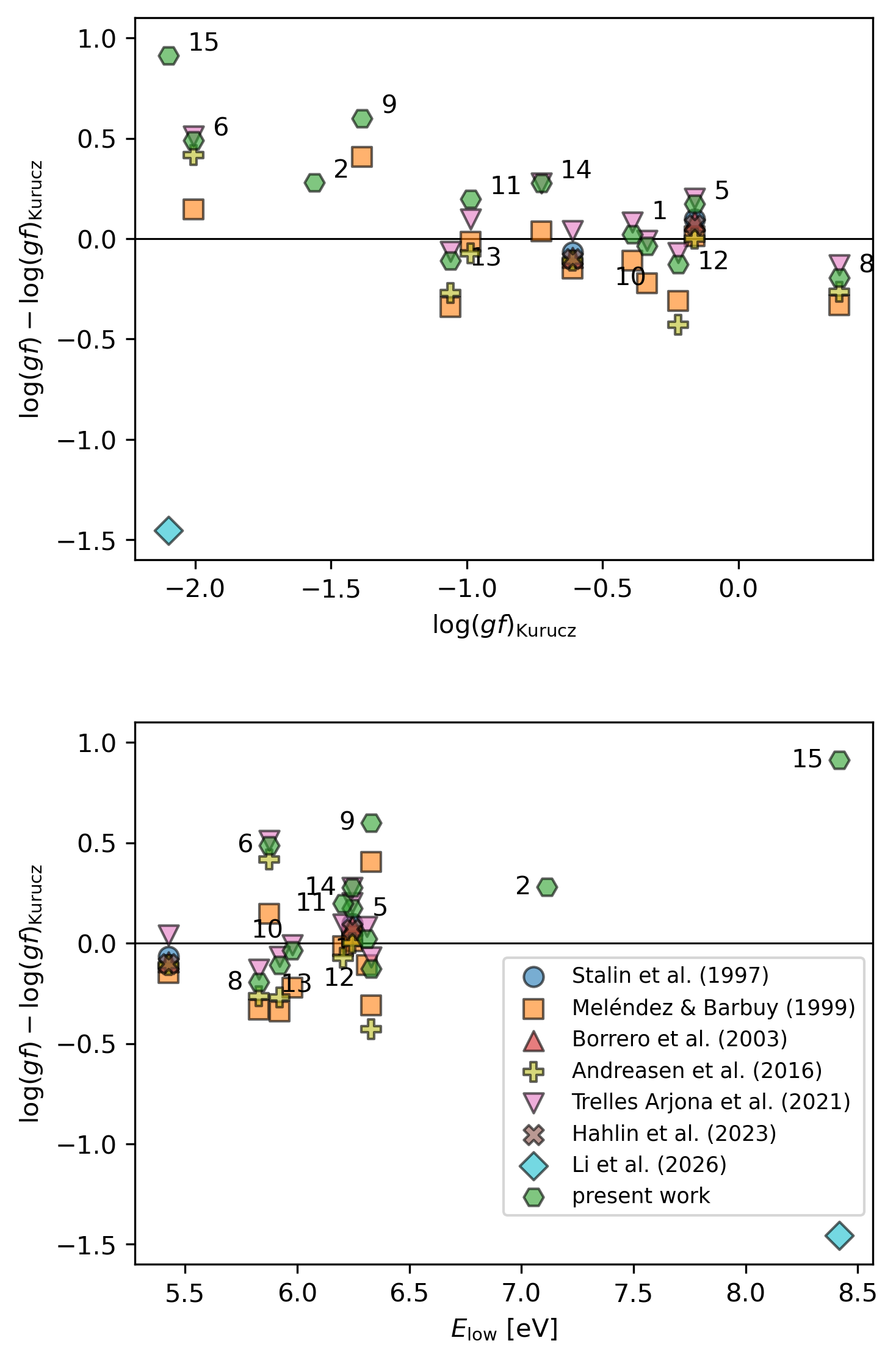}
    \caption{Differences between the \loggf{} values inferred from different studies and those found in the Kurucz database. \textit{Top panel:} \loggf{} differences with respect to \loggf{Kurucz}. \textit{Bottom panel:} \loggf{} differences with respect to the lower level excitation potential $E_\mathrm{low}$. Each data point from the present work is labelled by the corresponding line number from the Tab.\,\ref{tab:lines} in both panels.}
    \label{fig:loggf_compare}
\end{figure}

After establishing the robustness of the inferred \loggf{} values using the coupled method, in \fig{fig:loggf_compare} we compare them with the values determined by other authors using different methods and different solar observations. In the top panel, we contrast all sets of \loggf{} values against the widely used ones from the Kurucz database \citep{Kurucz95} \citep[identical \loggf{} values are also found in the VALD database,][]{Piskunov95}, while on the bottom panel we contrast them against the excitation potential of the lower level.

The \loggf{} values determined by different authors have small deviations compared to the values inferred with the coupled method. First, we will discuss the \loggf{} differences to the results from \cite{TrellesArjona21} since we use the same observations but different methods. 

One discrepancy between our work and \cite{TrellesArjona21} arises from differences in the adopted line lists. We started from the line list of \cite{TrellesArjona21} and obtained comparable \loggf{} values for all lines, albeit with a small systematic offset. However, we found that some of the lines have exceptionally large \loggf{} values, thus violating the Thomas–Reiche–Kuhn sum rule for line transition strengths. Once we enforced this physical constraint on the upper \loggf{} boundary, it indicated that additional transitions were needed to reproduce the observed spectra. We therefore curated the line list of \cite{TrellesArjona21} with transitions from the VALD database, retaining all lines whose line-core depth exceeded $0.1\%$ of the continuum intensity calculated using the FAL-C atmospheric model \citep{FALC}. However, lines that satisfy this criterion but have \loggf{} values higher than their true values, without violating the sum rule, cannot be identified a priori. One possible indication is that the inferred \loggf{} value reaches the imposed lower boundary, regardless of how low this boundary is set. For such lines, the upper limit of \loggf{} cannot be reliably constrained since their profiles are buried in the noise, as they are not prominent enough for the retrieval of the physical conditions in the solar atmosphere.

The \loggf{} values inferred with the coupled method are on average lower by 0.023 compared to those from \cite{TrellesArjona21} (for lines common to both line lists and after abundance correction). The reason behind this offset is most probably due to the degeneracy between temperature and \loggf{} values. For the particular set of spectral lines we considered here, there is a clear preference for larger \loggf{} values once the coupling between pixels is lifted (see \fig{fig:gradT_vs_loggf}). Increasing the temperature gradient in the deeper layers of the atmosphere steepens the source function, hence deepening the spectral lines, which leads to smaller inferred \loggf{} values. The severity of the temperature-\loggf{} degeneracy depends on the element to which the spectral line belongs and the excitation potential of the lower level of the transition. The most abundant lines in the observed wavelength range that we identified belong to the neutral iron atom, which are also the ones we used for comparing our results to those of \cite{TrellesArjona21}. Since all these lines also originate from higher excitation potential levels ($E_\mathrm{low}\geq5.4$\,eV), their temperature sensitivity is similar. This would then introduce a systematic offset in the inferred \loggf{} due to the temperature changes in the pixel-by-pixel inversions.

The stronger lines show much smaller deviations from Kurucz values, while the weaker ones exhibit larger deviations. The largest deviation is found for line 15, the weakest line in our line list. \cite{Li26} determined its \loggf{} from modelling the sulphur atom, where the line central wavelength is outside of the observed wavelength region. Thus, we could question whether this line is properly identified. Nevertheless, its effect on the nearby blend, line 14, is examined in the previous section in case V, where we found an increase of 5\% in \loggf{14} when line 15 is removed. Additionally, \cite{TrellesArjona21} inferred identical \loggf{14} even though another line was considered as a blend pair.

On the other hand, there is a small dependence of the \loggf{} deviations with respect to \loggf{Kurucz}, which points to some systematic effects. This trend is also recovered for the \loggf{} values from \cite{TrellesArjona21}, who used the same set of observations, and \cite{Andreasen16} and \cite{Melendez99}, who used a spatially averaged solar spectrum. Since \cite{TrellesArjona21} used a different method to determine the \loggf{} values, we can conclude that the trend is independent of the method used to infer \loggf{} values from the observations. Additionally, even using the different line lists does not affect the trend since differences between our results and those of \cite{TrellesArjona21} are smaller than the differences to the Kurucz values.

Since all three, \cite{TrellesArjona21}, \cite{Andreasen16}, and \cite{Melendez99}, used different observations and methods to determine \loggf{} values, we can assume that the recovered trend is real. Though, we have to be careful not to over-interpret it since the number of data points (lines) is not large. Unfortunately, we lack experimental measurements of \loggf{} values for these lines to firmly establish if the observed trend is real or a coincidence. Nevertheless, we will briefly discuss two other potential causes for this trend.

One of the causes for the trend in the \loggf{} difference with line strength could be found in NLTE effects. However, these lines are formed very deep in the solar atmosphere where frequent particle collisions decouple the level populations from the radiation field. If there were any NLTE effects, those would be evident from the bottom panel in \fig{fig:loggf_compare}. Since we lack any trend with the lower-level excitation potential, we can rule out the NLTE effects here. Furthermore, we would expect the NLTE effects to be smaller in magnitude compared to the deviations to the Kurucz values we observe here. \cite{Hahlin23} found that the NLTE corrections applied to some of the lines from the H-band spectrum altered the inferred \loggf{} values on the order of $\sim0.01$\,dex.

Based on our current experience with inferring the \loggf{} values with the coupled method from SCIP/Sunrise\textsc{III} observations at $850$\,nm, disregarding the NLTE effects would be compensated by altering the temperature gradient. Taking an example of the line originating from the same level as the reference line, the change in the temperature gradient caused its \loggf{} to decrease by $0.02$\,dex compared to the experimentally measured one. However, for a spectral line originating from a higher excitation potential level, which also has an experimentally measured \loggf{}, it decreased by around $0.07$\,dex. The strength of the NLTE effects would vary between lines of different ions and depend on the features we observe at the solar surface \citep{Smitha20}. For the lines we have analysed in the present study and the arguments provided above, the NLTE effects can be ruled out as the source of the trend from the top panel in \fig{fig:loggf_compare}.

Another explanation for large deviations from the Kurucz values could be found within the semi-empirical calculations of the \loggf{} values we are comparing with. The vast majority of lines we have considered here belong to atomic iron and originate from energy levels with lower excitation potential larger than $5.4$\,eV. These levels are poorly modelled in a simple spin-orbit interaction (LS coupling) framework due to overlaps in wave functions of higher-lying levels \citep[configuration interaction; e.g.][]{Clear26}. Thus, their modelling necessitates proper accounting for configuration interactions, which are hard to model completely and could result in \loggf{} errors of at least 0.2 for the LS-forbidden transitions \citep[e.g.][]{Li26}. The iron atom requires a very complex atomic model due to many open shells and densely packed higher-lying energy levels. As a community, we currently lack extensive modelling of the iron atom and theoretically derived transition probabilities \citep{Conroz18}. 

\section{Uncertainty estimate of \loggf{} values}
\label{sec:mcmc}

The Markov chain Monte Carlo \citep[MCMC;][]{Press07} provides a method to statistically evaluate the uncertainties for model parameters, where direct error propagation is impossible. The method is flexible enough and allows a full parameter exploration that guarantees the global minimum, but is very computationally demanding. Our primary goal here is to determine the uncertainties of the \loggf{} values.

For this purpose, we need to obtain the joint posterior distribution $p(\hat{L}, \hat{G}|D)$ for all $\hat{L}$ local (atmospheric) and $\hat{G}$ global (atomic) parameters for the given observations $D$. Applying Bayes' theorem, we can write the joint posterior as:
\begin{equation}
    p(\hat{L}, \hat{G}|D) \propto p(D|\hat{L},\hat{G})\cdot p(\hat{L})\cdot p(\hat{G}),
\end{equation}
where $p(D|\hat{L},\hat{G})$ is the likelihood function (proportional to $\exp(-\chi^2)$) describing how well the parameters represent the observations, and $p(\hat{L})$ and $p(\hat{G})$ are priors for the local and global parameters, respectively. 

The uncertainties for the parameters of interest are calculated as the standard deviation of the marginalized joint posterior distribution. In our case, we are primarily interested in obtaining the uncertainties for the \loggf{} parameter of all the lines in our line list. For that reason, we integrate the joint posterior distribution over the local parameters:
\begin{equation}
    p(\hat{G}|D) = \int p(\hat{L}, \hat{G}|D) \mathrm{d}\hat{L}.
\end{equation}

Following the principles of the coupled method, we can sample the local parameters from each pixel independently without any loss in generality. Thus, we decompose the joint posterior distribution as the product of the posterior distributions of all the pixels in our field of view:
\begin{equation}
    p(\hat{G}|D) = \prod_i \int p(\hat{L_i}, \hat{G}|D_i) \mathrm{d}\hat{L_i}.
\end{equation}
However, the global parameters are still assumed to be sampled simultaneously for all the pixels. To further simplify the sampling of the global parameters, we will allow them to vary between pixels and apply the data fusion framework \citep{Wu24} to combine the marginalized posteriors for global parameters from all the pixels. In this case, the marginalized posterior for the global parameters is:
\begin{equation}
    p(\hat{G}|D) = \prod_i \int p(\hat{L_i},\hat{G_i}|D_i) \mathrm{d}\hat{L_i} = \prod_i p(\hat{G_i}|D_i),
\end{equation}
where $p(\hat{G_i}|D_i)$ is the marginalized posterior for the global parameters for the $i$-th pixel. The data fusion allows us to properly combine the posteriors from different sources, groups, or measurements to improve the results for the common parameters. In our case, we can treat each pixel as an independent measurement and combine their posteriors to derive the joint marginalized posterior for the global parameters. 

We opted to use the conditionally independent posterior data fusion method \citep{Wu24} to approximate the true marginalized posterior for the global parameters. This approximation is shown to hold well if the prior distributions for the global parameters are non-informative \citep{Wu24}.

From all local parameters, we chose to sample only temperature, line of sight velocity and micro-turbulent velocity of the magnetized component, and magnetic filling factor. These parameters showed to have the strongest effect on the \loggf{} inference. The initial values were drawn from a tight ball around the reference inversion values. For all the parameters, we used the uniform prior distributions with the same boundaries that were specified during the reference inversion. In the case of the \loggf{} parameter, we opted for the boundaries of $\pm1$ from the reference inversion values.

We used the affine-invariant ensemble sampler \citep{GoodmanWeare10} implemented in the \texttt{emcee} Python module \citep{ForemanMackey13} to obtain the joint posterior distribution. The stepping through a parameter space used the differential evolution move \citep{Nelson14} with the default parameters. This move type allows us to sample posteriors that are not unimodal, which is expected for the atmospheric parameters due to the degeneracy between the parameters \citep[see for e.g.][]{inversion}.

For each drawn sample, we synthesized spectra with \globin{} and computed the $\chi^2$ in the same way as we did with the coupled method. This ensured that we were sampling the same joint posterior distribution as the coupled method sensed. However, for the purpose of determining the \loggf{} uncertainties, we used three times higher noise level, $\sigma=3.23\cdot10^{-3}$. We determined this value as the standard deviation of the difference between the observed and the best-fit spectra. In the ideal case, the standard deviation of this distribution should be equal to the noise in the data. However, because we approximate the true atmospheric conditions in the solar atmosphere, the differences are larger than ideal. The uncertainties caused by the atmospheric parametrisation are impossible to quantify from observations alone, as the exact physical conditions in the solar atmosphere are not known. One possible approach would be to repeat the analysis using synthetic spectra generated from a magnetohydrodynamic MHD model of the solar atmosphere, which is outside the scope of the present work.

Performing MCMC sampling for every pixel from the reference inversion is a prohibitively expensive task. Thus, we opted to sample a small number of pixels and restricted ourselves to the 10 pixels with the best fit (lowest $\chi^2$). Still, these pixels represent a variety of spectral line profiles that should resemble well the whole population from the reference case.

The samples were drawn until the total length of chains was at least 50 times larger than the autocorrelation time and burned at least 50\,\% of chains before any analysis. Because of the central limit theorem, the marginalized distributions were fitted with the normal distribution whose parameters are then used to perform data fusion across all 10 pixels. The \loggf{} uncertainties of all the lines were determined as the standard deviation of the fused marginalized posteriors (\FusionLG{}). These values are given in \tab{tab:lines}. 

For most of the lines, the uncertainties are on the order of $1$\,\% ($0.004$\,dex), except for lines 2, 6, and 9, which have higher uncertainties of $3-4$\,\%, while the largest uncertainty is retrieved for line 15 ($12$\,\%).

\begin{figure}
    \centering
    \includegraphics[width=0.85\linewidth]{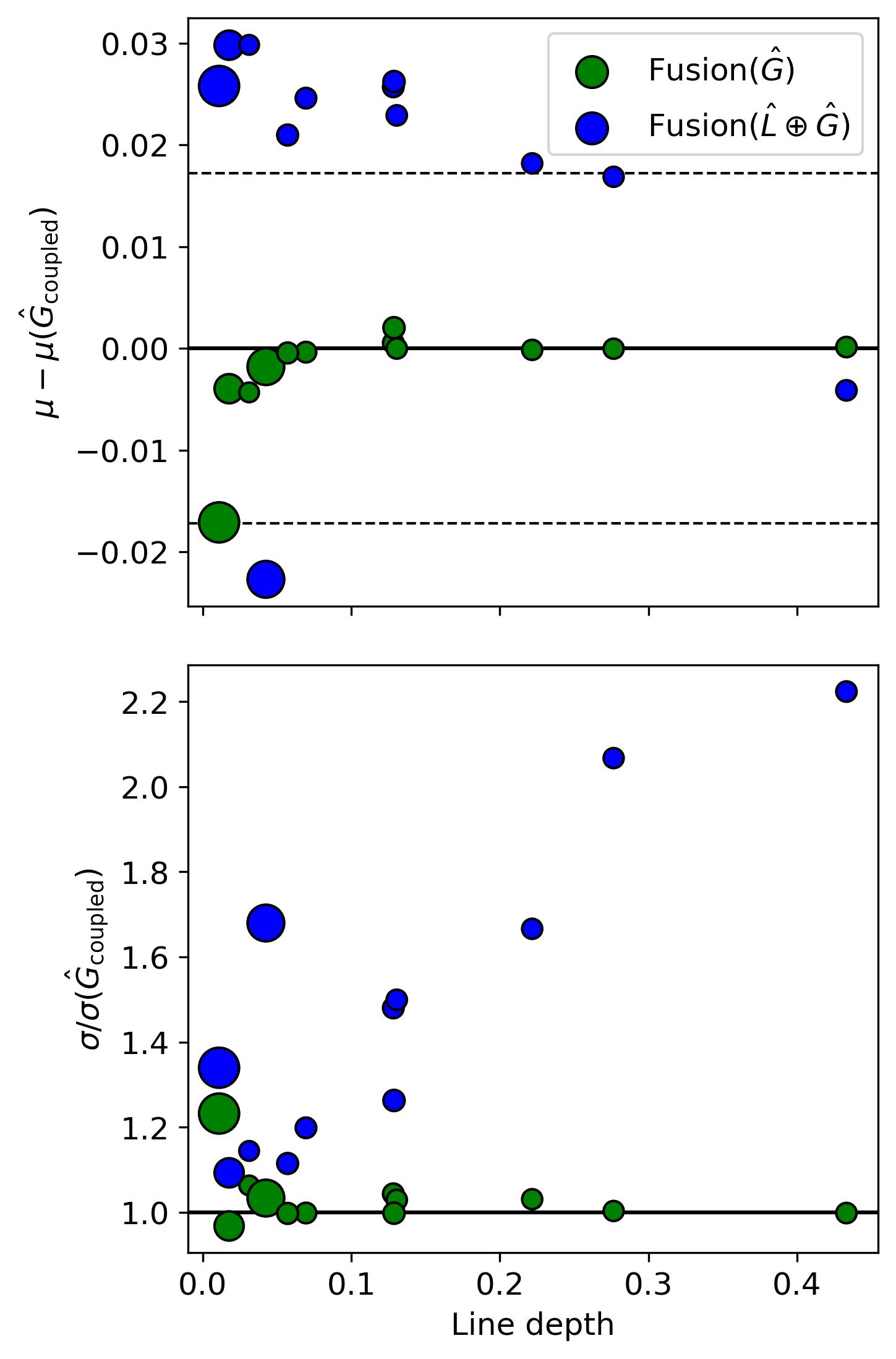}
    \caption{Comparison of the \loggf{} mean and standard deviation for fused MCMC models for the parameter spaces spanning atmospheric and atomic parameters (\FusionLG{}) and only atomic parameters (\FusionG{}), and the coupled atomic parameters (\Gcoupled). \textit{Top panel:} Differences between mean values for different parameter spaces. Dashed lines mark relative uncertainty of $4$\,\%. \textit{Bottom panel:} The ratio of the standard deviations between different parameter spaces. The symbol size in both panels represents how strongly the line is blended.}
    \label{fig:mcmc_std_mean_compare}
\end{figure}

To evaluate the validity of the data fusion process, we constructed two additional MCMC models. In the first one we sampled only the \loggf{} parameters pixel-by-pixel and then fused the samples across all 10 pixels (\FusionG{}). For this test, the atmospheric parameters were fixed to those from the reference inversion. In the second one, we sampled again only the \loggf{} values but simultaneously for the 10 best pixels (\Gcoupled{}). The comparison between the \loggf{} mean and standard deviation for each line between different MCMC models is given in \fig{fig:mcmc_std_mean_compare}.

The comparison between the mean \FusionG{} \loggf{} values and the \Gcoupled{} \loggf{} values is very well for all the lines, and the discrepancy is smaller than the reported uncertainties. This result validates the assumptions taken for the data fusion process and strengthens the accuracy of the determined \loggf{} uncertainties from the \FusionLG{} model.

Once the atmospheric parameters are sampled simultaneously with the \loggf{} parameter, the uncertainties increase. This increase differs between the lines and is larger for stronger lines. The weak lines are sensitive to changes in the atmospheric parameters at deeper layers, which are better constrained and thus have lower uncertainty (see \fig{fig:corner_atmospheric}). However, the stronger lines are formed at higher layers in the atmosphere where parameters are determined with larger uncertainty. Thus, it is expected that the \loggf{} uncertainty of stronger lines is larger than for the weaker ones. Once the spectral line is blended, this correlation is broken since the uncertainty in the \loggf{} values is shared between the blended lines. 

On the other hand, the offset in the mean \loggf{} values is smaller for the stronger lines, while the weaker lines are on average stronger by $\sim0.025$\,dex, which coincides well with the difference between the \loggf{} values inferred here and those from \cite{TrellesArjona21}. This difference is the consequence of the degeneracy between the temperature and \loggf{} mentioned before. The MCMC walkers locate the region in the parameter space that maximises the posterior probability for a given pixel. However, the \loggf{} values from one pixel do not necessarily correspond to the region with the maximum posterior in another pixel. Once the pixels are coupled together, the posterior maximum changes its location in the parameter space. The evidence of this could also be seen in Fig.\,4 and Fig.\,7 from \cite{Vukadinovic24}, where the temperature gradient is significantly different from the expected stratification and inferred \loggf{} values are, on average, larger by $0.025$\,dex. 

It is worth noting that we disregarded the uncertainty of the experimentally measured \loggf{} for line 4 and the uncertainties in the elemental abundances. These uncertainties are an order of magnitude larger than the determined \loggf{} uncertainties. To fully take these uncertainties into account, one could model these parameters assuming the Gaussian priors. Including the abundances in the model would further increase the degeneracy between all the parameters, and one would have to be careful in interpreting these results.

\section{Conclusions}
\label{sec:conclusions}

The reliability of atmospheric parameter determination from remote sensing observations of the solar atmosphere strongly depends on the quality of atomic data used to model the observed spectral lines. In the present work, we applied the coupled inversion method introduced in \cite{Vukadinovic24} to infer the \loggf{} values from the quiet-sun observations at $1.56$\,$\mu$m from the GRIS/GREGOR spectropolarimeter. 

In the coupled method, atomic parameters are the same for all spatial pixels (spatial coupling), whereas the atmospheric parameters are allowed to vary for every pixel individually. The absolute scale for \loggf{} values is established using the reference line whose \loggf{} value can be trusted. In the present work, we chose line 15648\,\AA{} as the reference, as it has an experimentally measured \loggf{} \citep{Ruffoni13}. The reference line is also used to reduce the degeneracy between the atmospheric and atomic parameters (e.g., between temperature and \loggf{}).

We tested and confirmed the robustness of the inferred \loggf{} values against uncorrected spectral stray light, dependence on the reference spectral line used to decouple the atmospheric and atomic parameters, selection bias of the field of view used to infer the \loggf{} values, and the influence of the chosen atmospheric model. All these tests suggest that spatial coupling in atomic line parameters ensures a robust and reliable \loggf{} inference. We also tested the impact of the macro-turbulent velocity on inferred \loggf{} values by lowering it from $1.56$\,km/s to $0.9$\,km/s. This change in the macro-turbulent velocity did not produce a meaningful change in the inversion results.

Additionally, we tested the effect of missing lines from the chosen line list. The missing blended lines caused an increase in the \loggf{} value of their blended pair, compensating for the lack of the opacity. However, the missing lines also produced an indirect effect on the \loggf{} values of other lines that are further away in wavelength. Due to the missing opacity, atmospheric parameters were adjusted, causing a decrease in the inferred \loggf{} values of other lines. This is most probably the consequence of having only the lines from a single element, in this case iron, which have the same temperature sensitivity.

The preceding results assume that the line list used for the inversions is complete and that all lines are adequately modelled under local thermodynamic equilibrium. Imposing physical constraints, such as the Thomas–Reiche–Kuhn sum rule, on the \loggf{} inference can further improve the reliability of the results and help identify misidentified lines or missing transitions in the adopted line list.

The uncertainties for the retrieved \loggf{} values were determined using the Markov Chain Monte Carlo method. The uncertainties for strong and weakly blended lines are on the order of $1$--$4$\,\%, whereas for strongly blended lines the uncertainty is $12$\,\%. These values are well matched with the uncertainties of experimentally measured \loggf{}'s \citep[1-5\,\%; ][]{Blackwell72} and to those from theoretical computations for strong lines \citep[5-10\,\%][]{Li21}. Based on these results, we can argue here that the \loggf{} values inferred from observed spectra using the coupled method have comparable uncertainty to other methods, and should be used if no better source of atomic data exists.

Applying the coupling in \loggf{} values, as done in the coupled method, the parameter space for \loggf{} values is constrained, and the parameter degeneracy is thus lifted. However, this is true as long as solar features with distinct and sufficiently different temperature stratifications are observed, which ensures the differential sensitivity of blended lines to atmospheric parameters (mainly temperature). The main strength of the coupled method is in removing the degeneracy between atomic and atmospheric parameters to retrieve \loggf{} values of blended lines accurately. The same principle can be applied to other atomic parameters as well.

The coupled method will be essential for the fitting of spectra with a large number of blended spectral lines, like the near-ultraviolet spectra obtained by the SUSI instrument \cite[]{Feller25} during the \textsc{Sunrise iii} flight in 2024 \cite[]{KorpiLagg25}. The high spatial resolution will additionally allow for the analysis of very different solar features, with more diverse temperature and magnetic structures, allowing for improvement not only of the \loggf{} determination, but also the determination of the physical conditions in the solar atmosphere.

\begin{acknowledgements}
We would like to thank the anonymous referee for their comments that significantly improved the presented results. DV would like to thank Ivan Mili\'{c} and Carlos Jos\'{e} Diaz Baso for fruitful discussions regarding the Bayesian inference. This work has made use of the VALD database, operated at Uppsala University, the Institute of Astronomy RAS in Moscow, and the University of Vienna. This project has received funding from the European Research Council (ERC) under the European Union's Horizon 2020 research and innovation programme (grant agreement No. 101097844 — project WINSUN, and grant agreement No. 101118581 - project REVEAL).
\end{acknowledgements}

\bibliography{references.bib}

\begin{appendix}
\nolinenumbers

\section{Atomic line parameters}

Table\,\ref{tab:lines} lists the atomic line parameters we used to model the observed spectra. We also list the \loggf{} values inferred by other researchers for comparison.

\begin{strip}
    \centering
    \captionof{table}{Atomic parameters used to model the observed spectral lines.}
    \begin{tabular}{c l c r c r c c c c c}
    \midrule[1.5pt]
    \multirow{2}{*}{No.} & \multirow{2}{*}{Element} & \multirow{2}{*}{$\lambda$ [\AA{}]} & \multirow{2}{*}{$E_l$ [eV]} & \multirow{2}{*}{$J_l$} & \multirow{2}{*}{$J_u$} & \multirow{2}{*}{$g_\mathrm{L}^\mathrm{eff}$} & \multicolumn{4}{c}{\loggf{}} \\
    &&&&&&& Kurucz & present work & TA21\tablefootmark{8} & Other \\ \midrule[1.5pt]
    1 & Fe I & 15645.016 & 6.3109 & 2.0 & 2.0 & 2.333 & -0.390 & -0.368 $\pm$ 0.004 & -0.308 $\pm$ 0.013 & -0.500\tablefootmark{5} \\ 
	2 & Si I & 15645.415 & 7.1133 & 2.0 & 3.0 & 1.086 & -1.561 & -1.281 $\pm$ 0.015 & ... & ... \\ 
	3 & Si I & 15645.843 & 7.1660 & 2.0 & 3.0 & 1.000 & -1.238 & ... & ... & ... \\ \midrule[1pt]
	4 & Fe I & 15648.510 & 5.4260 & 1.0 & 1.0 & 3.000 & -0.61\tablefootmark{6} & -0.61\tablefootmark{6} & -0.571 $\pm$ 0.008 & -0.705\tablefootmark{2} \\ &&&&&&&&&& -0.720\tablefootmark{1} \\ &&&&&&&&&& -0.712\tablefootmark{3} \\ &&&&&&&&&& -0.676\tablefootmark{7} \\ &&&&&&&&&& -0.760\tablefootmark{5} \\ \midrule[1pt]
	5 & Fe I & 15652.871 & 6.2461 & 5.0 & 4.0 & 1.670 & -0.161 & 0.013 $\pm$ 0.003 & 0.041 $\pm$ 0.010 & -0.073\tablefootmark{2} \\ &&&&&&&&&& -0.160\tablefootmark{1} \\ &&&&&&&&&& -0.090\tablefootmark{3} \\ &&&&&&&&&& -0.063\tablefootmark{7} \\ &&&&&&&&&& -0.150\tablefootmark{5} \\ \midrule[1pt]
	6 & Fe I & 15656.632 & 5.8741 & 4.0 & 3.0 & 1.875 & -2.005 & -1.517 $\pm$ 0.011 & -1.494 $\pm$ 0.044 & -1.598\tablefootmark{1} \\ &&&&&&&&&& -1.860\tablefootmark{5} \\ 
	7 & Fe I & 15656.767 & 6.4258 & 1.0 & 2.0 & 2.000 & -1.873 & ... & ... & ... \\ \midrule[1pt]
	8 & Fe I & 15662.013 & 5.8279 & 5.0 & 4.0 & 1.500 & 0.371 & 0.176 $\pm$ 0.003 & 0.240 $\pm$ 0.013 & 0.097\tablefootmark{1} \\ &&&&&&&&&& 0.040\tablefootmark{5} \\ 
	9 & Fe I & 15662.313 & 6.3292 & 1.0 & 2.0 & 0.111 & -1.386 & -0.786 $\pm$ 0.014 & ... & -0.980\tablefootmark{5}\\ \midrule[1pt]
	10 & Fe I & 15665.240 & 5.9787 & 1.0 & 1.0 & 0.750 & -0.337 & -0.374 $\pm$ 0.002 & -0.346 $\pm$ 0.009 & -0.560\tablefootmark{5} \\ \midrule[1pt]
	11 & Fe I & 15670.124 & 6.2057 & 5.0 & 5.0 & 1.550 & -0.986 & -0.789 $\pm$ 0.005 & -0.890 $\pm$ 0.026 & -1.060\tablefootmark{1} \\ &&&&&&&&&& -1.000\tablefootmark{5} \\ \midrule[1pt]
	12 & Fe I & 15671.004 & 6.3292 & 1.0 & 2.0 & 0.424 & -0.221 & -0.348 $\pm$ 0.003 & -0.290 $\pm$ 0.016 & -0.650\tablefootmark{1} \\ &&&&&&&&&& -0.530\tablefootmark{5} \\ \midrule[1pt]
	13 & Fe I & 15671.866 & 5.9209 & 3.0 & 3.0 & 1.083 & -1.060 & -1.169 $\pm$ 0.006 & -1.125 $\pm$ 0.030 & -1.330\tablefootmark{1} \\ &&&&&&&&&& -1.400\tablefootmark{5} \\ \midrule[1pt]
	14 & Fe I & 15673.151 & 6.2461 & 5.0 & 6.0 & 0.564 & -0.726 & -0.450 $\pm$ 0.005 & -0.448 $\pm$ 0.020 & -0.690\tablefootmark{5} \\
	15 & S I & 15673.308 & 8.4173 & 3.0 & 2.0 & 1.500 & -2.098 & -1.185 $\pm$ 0.048 & ... & -3.553\tablefootmark{4} \\ 
    \midrule[1.5pt]
    \end{tabular}
    \tablefoot{The spectral lines are sorted by ascending wavelength. The columns represent the line number in the analysed spectral region, the chemical element and ionisation stage, the central wavelength of the line, the energy of the lower state involved in the transition, the total angular momentum of the lower and upper levels, the effective Land\'e factor, and \loggf{} values from the Kurucz database, values with uncertainty determined in the present work, those from \cite{TrellesArjona21}, and from other sources (corrected for abundance differences where applicable). All atomic data are from the Kurucz database \citep{Kurucz95}. Elemental abundances are taken from \cite{Asplund21}, and are $A_\mathrm{Fe}=7.46$, $A_\mathrm{Si}=7.51$, and $A_\mathrm{S}=7.12$. The uncertainties in the abundances were not propagated into the \loggf{} uncertainties. Different blend pairs are separated by horizontal lines.\\
    References.
    \tablefoottext{1}{\cite{Andreasen16}}
    \tablefoottext{2}{\cite{Borrero03}}
    \tablefoottext{3}{\cite{Hahlin23}}
    \tablefoottext{4}{\cite{Li26}}
    \tablefoottext{5}{\cite{Melendez99}}
    \tablefoottext{6}{\cite{Ruffoni13}}
    \tablefoottext{7}{\cite{Stalin97}}
    \tablefoottext{8}{\cite{TrellesArjona21}}
    }
    \label{tab:lines}
\end{strip}

\section{MCMC results}
\label{sec:MCMC_best_10}

\begin{figure}[h!]
    \centering
    \includegraphics[width=\linewidth]{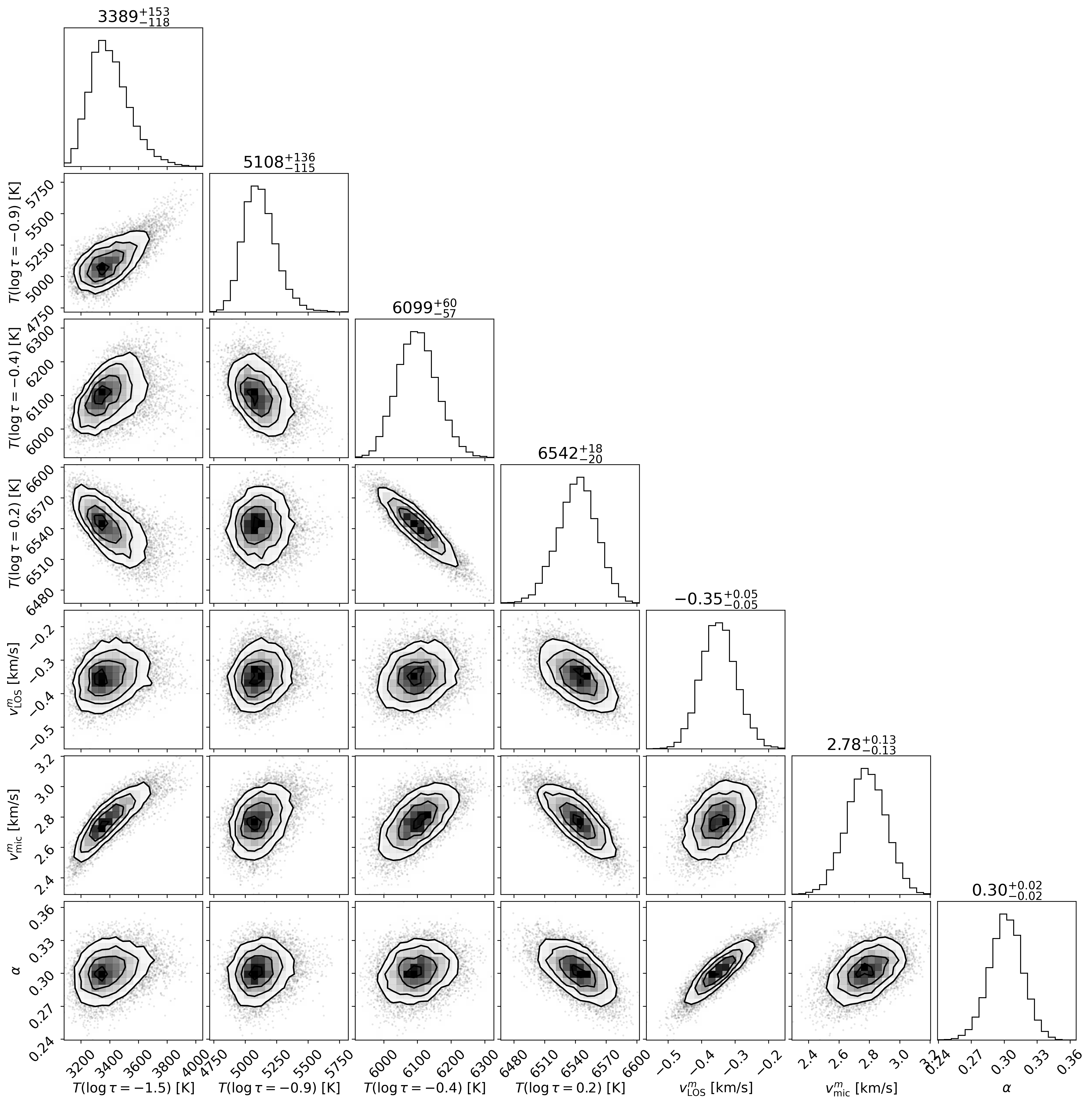}
    \caption{Corner plot of the temperature, line of sight velocity and micro-turbulent velocity of the magnetised component, and the magnetic filling factor for the pixel with the best fit from the reference inversion. The values at the top of the first panels in each column show the mean values with lower and upper uncertainties computed as $16\%$ and $84\%$ quantiles, respectively.}
    \label{fig:corner_atmospheric}
\end{figure}

In Figure\,\ref{fig:corner_atmospheric} we show a corner plot for the temperature, line of sight velocity and micro-turbulent velocity of the magnetised component, and stray light factor for the pixel with the best fit. Here we observe a correlation between certain parameters. For example, there is a negative correlation between the bottom two temperature nodes. Increasing the temperature at the base of the atmosphere (at $\log\tau_{500}=0.2$) will cause a decrease in the temperature node above it (at $\log\tau_{500}=-0.4$). This is the consequence of temperature interpolation and fixing the temperature gradient at the bottom node to the temperature gradient in the FAL C atmospheric model.

Another correlation is observed between temperature and microturbulent velocity, as both parameters contribute to spectral line broadening. In the lower atmosphere, a decrease in temperature is compensated for by an increase in microturbulent velocity, resulting in an anticorrelation between the two parameters that helps preserve the line widths. In the upper atmosphere, however, the parameters are positively correlated, with increases in both temperature and microturbulent velocity contributing to broader spectral lines.

We find that the temperature in the lower atmosphere is the least uncertain, with uncertainty increasing at higher layers. The temperature in the lower atmosphere is well constrained by the continuum, which provides numerous wavelength points. In contrast, the temperature in the upper atmosphere is primarily constrained by the cores of strong spectral lines. Since relatively few lines probe these higher layers, the constraints on the temperature become weaker, resulting in a broader range of permissible parameter values.

\begin{figure}[h!]
    \centering
    \includegraphics[width=0.9\linewidth]{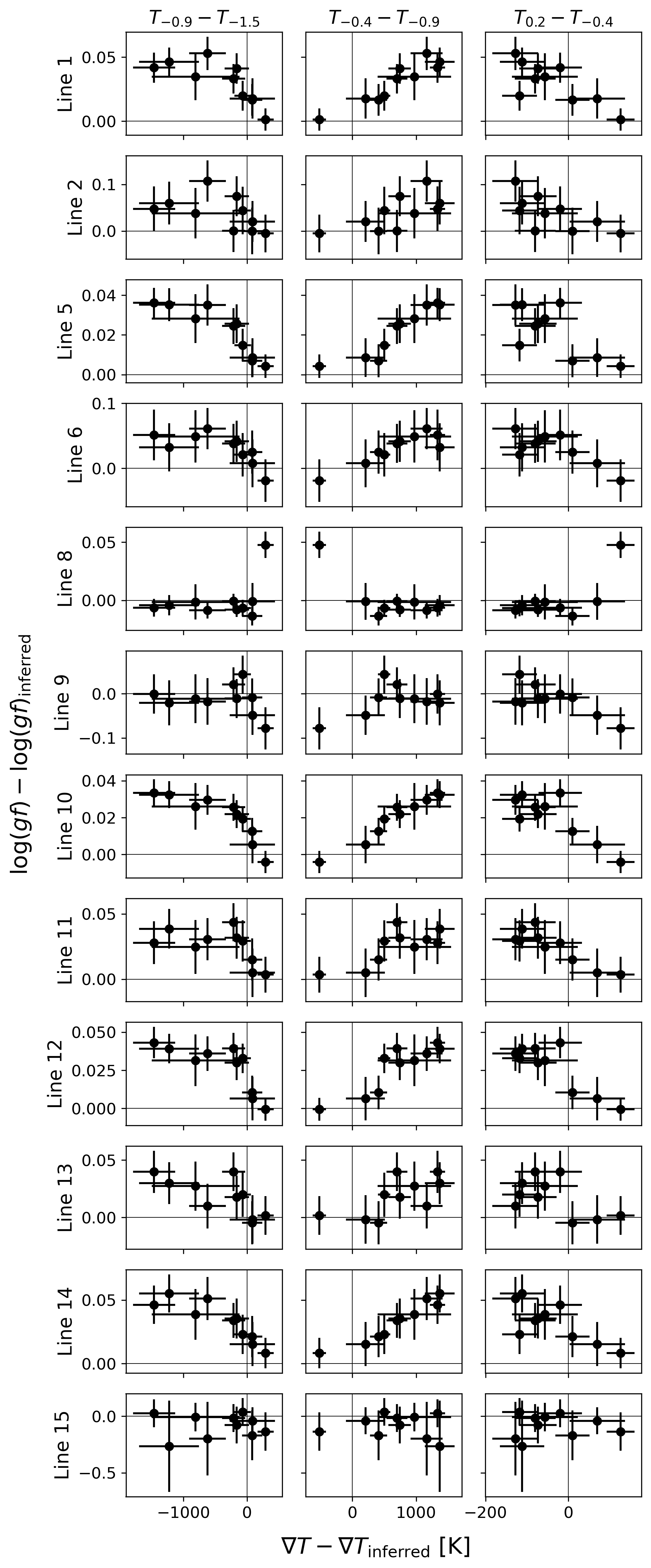}
    \caption{\loggf{} differences versus the differences in the temperature gradient. The x-axis displays the difference between the temperature gradient inferred from the MCMC and the coupled method, where the y-axis shows the difference between the \loggf{} inferred from the MCMC and the coupled method. The temperature gradient is computed between consecutive nodes whose height in $\log\tau_{500}$ is indicated as the index in the x-axis labels. Each row represents a different spectral line from Tab.\,\ref{tab:lines}.}
    \label{fig:gradT_vs_loggf}
\end{figure}

The effect of the pixel-by-pixel inference on the retrieved \loggf{} values is shown in Figure\,\ref{fig:gradT_vs_loggf} where we combined MCMC results from all 10 pixels we analysed. In this figure, we show the deviations in the temperature gradient and \loggf{} obtained in the MCMC sampling compared to the results from the reference inversion in Section\,\ref{sec:inv_patch_A}. In virtually all the lines, we observe that \loggf{} increases with the decreasing temperature gradient at the bottom of the atmosphere. This is a consequence of the aforementioned correlation between temperature in successive nodes. We observe here that the correlation alternates as we observe the temperature gradient between successive nodes.

\end{appendix}

\end{document}